\documentclass[twocolumn]{aastex701}
\usepackage{graphicx}
\usepackage{multirow}
\usepackage{caption}
\usepackage{subcaption}
\usepackage{xcolor}

\usepackage{soulutf8}

\begin{document}

\title{The ALMA-QUARKS Survey. IV. Core Mass Function and Evolutionary Trends in Core Formation Efficiency and Chemistry}

\author[orcid=0000-0001-9822-7817,gname=Wenyu,sname=Jiao]{Wenyu Jiao}
\affiliation{State Key Laboratory of Radio Astronomy and Technology, Shanghai Astronomical Observatory, Chinese Academy of Sciences, 80 Nandan Road, Shanghai 200030, People's Republic of China}
\email[show]{astrojiao@gmail.com}

\author[orcid=0000-0002-5286-2564,gname=Tie,sname=Liu]{Tie Liu}
\affiliation{State Key Laboratory of Radio Astronomy and Technology, Shanghai Astronomical Observatory, Chinese Academy of Sciences, 80 Nandan Road, Shanghai 200030, People's Republic of China}
\email[show]{liutie@shao.ac.cn}

\author[orcid=0009-0004-6159-5375,gname=Dongting,sname=Yang]{Dongting Yang}
\affiliation{State Key Laboratory of Radio Astronomy and Technology, Shanghai Astronomical Observatory, Chinese Academy of Sciences, 80 Nandan Road, Shanghai 200030, People's Republic of China}
\email{dongting@shao.ac.cn}

\author[orcid=0000-0001-5950-1932,gname=Fengwei,sname=Xu]{Fengwei Xu}
\affiliation{Max-Planck-Institut f\"ur Astronomie, K\"onigstuhl 17, D-69117 Heidelberg, Germany}
\email{fengweilookuper@gmail.com}

\author[orcid=0000-0003-1649-7958,gname=Guido,sname=Garay]{Guido Garay}
\affiliation{Chinese Academy of Sciences South America Center for Astronomy, National Astronomical Observatories, Chinese Academy of Sciences, Beijing 100101, People's Republic of China}
\affiliation{Departamento de Astronom\'ia, Universidad de Chile, Camino el Observatorio 1515, Las Condes, Santiago 7591245, Chile}
\email{guido@das.uchile.cl}

\author[orcid=0000-0002-7125-7685,gname=Patricio,sname=Sanhueza]{Patricio Sanhueza}
\affiliation{Department of Astronomy, School of Science, The University of Tokyo, 7-3-1 Hongo, Bunkyo-ku, Tokyo 113-0033, Japan}
\email{patosanhueza@gmail.com}

\author[orcid=0000-0003-2412-7092,gname=Kee-Tae,sname=Kim]{Kee-Tae Kim}
\affiliation{Korea Astronomy and Space Science Institute, 776 Daedeokdae-ro, Yuseong-gu, Daejeon 34055, Republic of Korea}
\affiliation{University of Science and Technology (UST), 217 Gajeong-ro, Yuseong-gu, Daejeon 34113, Republic of Korea}
\email{ktkim@kasi.re.kr}

\author[orcid=0000-0001-7817-1975,gname=Yankun,sname=Zhang]{Yankun Zhang}
\affiliation{State Key Laboratory of Radio Astronomy and Technology, Shanghai Astronomical Observatory, Chinese Academy of Sciences, 80 Nandan Road, Shanghai 200030, People's Republic of China}
\email{zhangyankun@shao.ac.cn}

\author[orcid=0000-0002-9836-0279,gname=Siju,sname=Zhang]{Siju Zhang}
\affiliation{Departamento de Astronom\'ia, Universidad de Chile, Camino el Observatorio 1515, Las Condes, Santiago 7591245, Chile}
\email{sijuzhangastro@gmail.com}

\author[orcid=0000-0001-8315-4248,gname=Xunchuan,sname=Liu]{Xunchuan Liu}
\affiliation{Leiden Observatory, Leiden University, PO Box 9513, 2300 RA Leiden, The Netherlands}
\email{liuxunchuan001@gmail.com}

\author[gname=Lei,sname=Zhu]{Lei Zhu}
\affiliation{Chinese Academy of Sciences South America Center for Astronomy, National Astronomical Observatories, Chinese Academy of Sciences, Beijing 100101, People's Republic of China}
\email{lzhupku@gmail.com}

\author[orcid=0000-0003-2300-8200,gname=Amelia M.,sname=Stutz]{Amelia M. Stutz}
\affiliation{Departamento de Astronom\'ia, Universidad de Concepci\'on, Concepci\'on, Chile}
\email{amelia.stutz@gmail.com}

\author[orcid=0000-0002-5809-4834,gname=Mika,sname=Juvela]{Mika Juvela}
\affiliation{Department of Physics, University of Helsinki, PO Box 64, FI-00014 Helsinki, Finland}
\email{mika.juvela@helsinki.fi}

\author[orcid=0000-0001-5175-1777,gname=Neal J.,sname=Evans]{Neal J. Evans}
\affiliation{Department of Astronomy, The University of Texas at Austin, 2515 Speedway, Austin, TX 78712, USA}
\email{nje@astro.as.utexas.edu}

\author[gname=Shanghuo,sname=Li]{Shanghuo Li}
\affiliation{School of Astronomy and Space Science, Nanjing University, 163 Xianlin Avenue, Nanjing 210023, People's Republic of China}
\affiliation{Key Laboratory of Modern Astronomy and Astrophysics (Nanjing University), Ministry of Education, Nanjing 210023, People's Republic of China}
\email{shli@nju.edu.cn}

\author[orcid=0000-0002-8697-9808,gname=Sami,sname=Dib]{Sami Dib}
\affiliation{Max-Planck-Institut f\"ur Astronomie, K\"onigstuhl 17, D-69117 Heidelberg, Germany}
\email{sami.dib@gmail.com}

\author[orcid=0000-0002-6622-8396,gname=Paul F.,sname=Goldsmith]{Paul F. Goldsmith}
\affiliation{Jet Propulsion Laboratory, California Institute of Technology, Pasadena, CA 91109, USA}
\email{paul.f.goldsmith@jpl.nasa.gov}

\author[orcid=0000-0003-4546-2623,gname=Aiyuan,sname=Yang]{Aiyuan Yang}
\affiliation{National Astronomical Observatories, Chinese Academy of Sciences, Beijing, 100101, People's Republic of China}
\affiliation{Key Laboratory of Radio Astronomy and Technology, Chinese Academy of Sciences, A20 Datun Road, Chaoyang District, Beijing, 100101, People's Republic of China}
\email{yangay@bao.ac.cn}

\author[orcid=0000-0001-7866-2686,gname=Jihye,sname=Hwang]{Jihye Hwang}
\affiliation{Institute for Advanced Study, Kyushu University, 774 Motooka, Nishi-ku, Fukuoka 819-0395, Japan}
\affiliation{Department of Earth and Planetary Sciences, Faculty of Science, Kyushu University, 774 Motooka, Nishi-ku, Fukuoka 819-0395, Japan}
\email{hwang.jihye.514@m.kyushu-u.ac.jp}

\author[orcid=0000-0003-3010-7661,gname=Di,sname=Li]{Di Li}
\affiliation{New Cornerstone Science Laboratory, Department of Astronomy, Tsinghua University, Beijing 100084, China}
\affiliation{Radio Astronomy National Key Laboratory, National Astronomical Observatories, Chinese Academy of Sciences, Beijing 100101, China}
\email{dili@mail.tsinghua.edu.cn}

\author[orcid=0009-0002-9273-7220,gname=Gyuho,sname=Lee]{Gyuho Lee}
\affiliation{Korea Astronomy and Space Science Institute, 776 Daedeokdae-ro, Yuseong-gu, Daejeon 34055, Republic of Korea}
\affiliation{University of Science and Technology (UST), 217 Gajeong-ro, Yuseong-gu, Daejeon 34113, Republic of Korea}
\email{space295@kasi.re.kr}

\author[orcid=0000-0002-4154-4309,gname=Xindi,sname=Tang]{Xindi Tang}
\affiliation{Xinjiang Astronomical Observatory, Chinese Academy of Sciences, Urumqi 830011, People's Republic of China}
\email{tangxindi@xao.ac.cn}

\author[orcid=0000-0002-3179-6334,gname=Chang Won,sname=Lee]{Chang Won Lee}
\affiliation{Korea Astronomy and Space Science Institute, 776 Daedeokdae-ro, Yuseong-gu, Daejeon 34055, Republic of Korea}
\email{cwl@kasi.re.kr}

\author[orcid=0000-0002-9793-3039,gname=Anindya,sname=Saha]{Anindya Saha}
\affiliation{Kavli Institute for Astronomy and Astrophysics, Peking University, 5 Yiheyuan Road, Haidian District, Beijing 100871, People's Republic of China}
\email{anindya.s1130@gmail.com}

\author[orcid=0000-0001-8812-8460,gname=Naval Kishor,sname=Bhadari]{Naval Kishor Bhadari}
\affiliation{Kavli Institute for Astronomy and Astrophysics, Peking University, 5 Yiheyuan Road, Haidian District, Beijing 100871, People's Republic of China}
\email{naval1996kishor@gmail.com}

\author[orcid=0009-0000-0178-7472]{Yuhan Yang}
\affiliation{Shanghai Astronomical Observatory, Chinese Academy of Sciences, No.80 Nandan Road, Shanghai 200030, People's Republic of China}
\affiliation{School of Astronomy and Space Sciences, University of Chinese Academy of Sciences,\\
No.19A Yuquan Road, Beijing 100049, People's Republic of China}
\email{yangyuhan@shao.ac.cn}

\author[orcid=0000-0001-7151-0882,gname=Swagat R.,sname=Das]{Swagat R. Das}
\affiliation{Departamento de Astronom\'ia, Universidad de Chile, Camino el Observatorio 1515, Las Condes, Santiago 7591245, Chile}
\email{swagat@das.uchile.cl}

\author[gname=Dmitry A.,sname=Ladeyschikov]{Dmitry A. Ladeyschikov}
\affiliation{Ural Federal University, Yekaterinburg 620062, Russia}
\email{dmitry.ladeyschikov@urfu.ru}

\author[gname=Shivani,sname=Gupta]{Shivani Gupta}
\affiliation{Indian Institute of Astrophysics, Koramangala II Block, Bangalore 560034, India}
\affiliation{Pondicherry University, R.V. Nagar, Kalapet, Puducherry 605014, India}
\affiliation{S. N. Bose National Centre for Basic Sciences, Block JD, Sector III, Salt Lake, Kolkata 700106, India}
\email{shivani.gupta@iiap.res.in}

\author{Sheng-li Qin}
\affiliation{School of Physics and Astronomy, Yunnan University, Kunming 650091, People’s Republic of China}
\email{qin@ynu.edu.cn}

\author{Hong-li Liu}
\affiliation{School of Physics and Astronomy, Yunnan University, Kunming 650091, People’s Republic of China}
\email{hongliliu2012@gmail.com}

\author[gname=Xiaofeng,sname=Mai]{Xiaofeng Mai}
\affiliation{State Key Laboratory of Radio Astronomy and Technology, Shanghai Astronomical Observatory, Chinese Academy of Sciences, 80 Nandan Road, Shanghai 200030, People's Republic of China}
\email{maixf@shao.ac.cn}

\author[orcid=0000-0003-1602-6849,gname=Prasanta,sname=Gorai]{Prasanta Gorai}
\affiliation{Center for Astronomy of Heidelberg University, Institute of Theoretical Astrophysics, Albert-Ueberle-Str.~2, 69120 Heidelberg, Germany}
\email{prasanta.astro@gmail.com}

\author[orcid=0000-0003-0295-6586,gname=Tapas,sname=Baug]{Tapas Baug}
\affiliation{S. N. Bose National Centre for Basic Sciences, Block JD, Sector III, Salt Lake, Kolkata 700106, India}
\email{tapas.polo@gmail.com}

\author[0000-0002-9875-7436]{James O. Chibueze}
\affiliation{UNISA Centre for Astrophysics and Space Sciences (UCASS),
College of Science, Engineering and Technology, University of South Africa, Cnr Christian de Wet Rd and Pioneer Avenue, Florida Park, 1709, Roodepoort, South Africa}
\affiliation{Department of Physics and Astronomy, Faculty of Physical Sciences, University of Nigeria, Carver Building, 1 University Road, Nsukka 410001, Nigeria}
\email{james.chibueze@gmail.com}

\author[orcid=0000-0002-8586-6721,gname=Pablo,sname=Garc\'ia]{Pablo Garc\'ia}
\affiliation{Chinese Academy of Sciences South America Center for Astronomy, National Astronomical Observatories, Chinese Academy of Sciences, Beijing 100101, People's Republic of China}
\affiliation{Instituto de Astronom\'ia, Universidad Cat\'olica del Norte, Av. Angamos 0610, Antofagasta, Chile}
\email{astro.pablo.garcia@gmail.com}

\author[orcid=0000-0002-9574-8454,gname=Leonardo,sname=Bronfman]{Leonardo Bronfman}
\affiliation{Departamento de Astronom\'ia, Universidad de Chile, Camino el Observatorio 1515, Las Condes, Santiago 7591245, Chile}
\email{leo@das.uchile.cl}

\author[gname=Andrej M.,sname=Sobolev]{Andrej M. Sobolev}
\affiliation{Xinjiang Astronomical Observatory, Chinese Academy of Sciences, Urumqi 830011, People's Republic of China}
\email{andrej.sobolev@urfu.ru}

\author[gname=Jian-Wen,sname=Zhou]{Jian-Wen Zhou}
\affiliation{Max-Planck-Institut f\"ur Radioastronomie, Auf dem H\"ugel 69, D-53121 Bonn, Germany}
\email{a18893153117@163.com}

\author[gname=Yunfan,sname=Jiao]{Yunfan Jiao}
\affiliation{Shanghai Astronomical Observatory, Chinese Academy of Sciences, No.80 Nandan Road, Shanghai 200030, People's Republic of China}
\affiliation{School of Astronomy and Space Sciences, University of Chinese Academy of Sciences,\\
No.19A Yuquan Road, Beijing 100049, People's Republic of China}
\email{yunfan.astro@gmail.com}

\author[gname=Jianjun,sname=Zhou]{Jianjun Zhou}
\affiliation{Xinjiang Key Laboratory of Radio Astrophysics, Urumqi 830011, People's Republic of China}
\email{zhoujj@xao.ac.cn}

\author[0009-0009-8154-4205]{Li Chen}
\affiliation{School of Physics and Astronomy, Yunnan University, Kunming 650091, People’s Republic of China}
\email{li.chen@mail.ynu.edu.cn}

\author[0000-0001-5710-6509]{Chuanshou Li}
\affiliation{School of Physics and Astronomy, Yunnan University, Kunming 650091, People’s Republic of China}
\email{lichuanshou2021@163.com}

\author[0009-0005-4295-5010]{Zhenying Zhang}
\affiliation{School of Physics and Astronomy, Yunnan University, Kunming 650091, People’s Republic of China}
\email{zyzhang.astro@gmail.com}

\author[gname=Jixiang,sname=Weng]{Jixiang Weng}
\affiliation{Shanghai Astronomical Observatory, Chinese Academy of Sciences, No.80 Nandan Road, Shanghai 200030, People's Republic of China}
\affiliation{School of Astronomy and Space Sciences, University of Chinese Academy of Sciences,\\
No.19A Yuquan Road, Beijing 100049, People's Republic of China}
\email{wengjixiang@shao.ac.cn}

\author[orcid=0000-0002-5310-4212,gname=L. Viktor,sname=T\'oth]{L. Viktor T\'oth}
\affiliation{Institute of Physics and Astronomy, E\"otv\"os Lor\'and University, P\'azm\'any P\'eter s\'et\'any 1/A, H-1117 Budapest, Hungary}
\affiliation{Faculty of Science and Technology, University of Debrecen, H-4032 Debrecen, Hungary}
\email{toth.laszlo.viktor@ttk.elte.hu}

\begin{abstract}

The ALMA-QUARKS (Querying Underlying Mechanisms of Massive Star Formation with ALMA-Resolved Gas Kinematics and Structures) survey targets 139 massive protoclusters selected from the ATOMS survey. The targets were observed at 1.3~mm wavelength with an angular resolution of $\sim0.3^{\prime\prime}$, corresponding to $\sim10^3$~au at the sample's median distance of $\sim3.8$~kpc. Using \texttt{getsf}, we identify 3374 dense cores and classify them as starless, protostellar, hot core, or H{\sc ii} region candidates based on spectral-line diagnostics (e.g., H30$\alpha$, CH$_3$CN, and CH$_3$OH) and CO outflow associations. The resulting catalog provides core physical properties and enables statistical investigations of dense core populations in massive protocluster environments. The derived global core mass function (CMF) after excluding H{\sc ii} region candidates, which is biased by distance effects, has a relatively flat power-law slope at the high-mass end ($\mathrm{d}N/\mathrm{d}\log M \propto M^{-\gamma}$, with $\gamma=0.98\pm0.03$). In contrast, the measured high-mass slopes in individual distance bins are approximately Salpeter-like, while their fitted lower mass limits increase with distance. We further examine the core formation efficiency (CFE) across the full sample and find that it positively correlates with the luminosity-to-mass ratio of the parental clumps, consistent with an evolutionary increase in the fraction of mass concentrated into dense cores. We also investigate the detection rates of different molecular tracers and find systematic chemical variations with evolutionary stage. Compared with the ASHES survey, we find that the detection rates of deuterated species vary significantly even for cores at the same evolutionary stages, suggesting an important environmental influence on the chemical evolution of cores.

\end{abstract}



\section{Background} 

High-mass stars ($M_*>8~M_{\odot}$) play a key role in galaxy evolution through their strong radiation, powerful stellar winds, and eventual supernova explosions, which inject energy and enrich the interstellar medium. However, the physical processes of high-mass star formation are still poorly constrained \citep{Tan_2014,Motte_2018a}. In contrast to low-mass stars, which can form in relatively isolated environments, high-mass stars typically form within massive, dense, cluster-forming regions \citep{Garay_1999,Lada_2003,Zinnecker_2007}. Two broad theoretical frameworks have been proposed for the assembly of high-mass stars: models in which the stellar mass reservoir is largely established in initially massive prestellar cores (core-fed scenario; \citealt{McKee_2003}), and models in which cores and protostars continue to gain mass from gas supplied by their surrounding clumps (clump-fed scenario; \citealt{Bonnell_2001,Bonnell_2004,Vazquez-Semadeni_2019,Padoan_2020}).

A key observational distinction between these frameworks is whether massive prestellar cores exist. Over the past decades, infrared dark clouds (IRDCs), widely believed to represent the cold and dense initial conditions of massive star formation, have therefore become prime targets in the search for such objects (e.g., \citealt{Egan_1998,Rathborne_2006,Simon_2006,Peretto_2009}). In recent years, interferometric surveys have resolved the internal structure of IRDCs down to (sub-)arcsecond scales, revealing extensive fragmentation and rich populations of dense cores (e.g., \citealp{Rathborne_2007, Zhang_2009, Wang_2011, Sanhueza_2017, Sanhueza_2019_ASHES, Svoboda_2019, Li_2019, Li_2020b, Morii2021, Morii2023}). Despite these advances, only a small number of high-mass prestellar core candidates have been identified (e.g., \citealp{Duarte_2013, Wang_2014, Barnes_2023, Mai_2024, Yang_2025}). Moreover, several of these candidates are located in regions that already exhibit active star formation rather than in the most quiescent IRDCs. The apparent scarcity of such objects in the earliest phases, together with observations of widespread infall and continuous gas feeding from clump to core scales, suggests that dense cores may continue to gain mass as their parental clumps evolve (e.g., \citealt{Peretto_2013, Pillai2023, Kong_2021, Xu_2023, Morii2026}).

Understanding how dense cores assemble and evolve is therefore critical for uncovering the origin of high-mass stars \citep{Beuther_2025}. Observational studies over the past decade have provided detailed views of dense cores in massive star-forming regions (e.g., \citealt{Palau_2013,Ginsburg_2018,Cao_2019,Nony_2023,Li_2024,Li_2025,Coletta_2025_ALMAGAL}). Some studies report fragmentation levels consistent with thermal Jeans predictions, suggesting that gravity and thermal pressure dominate on core-forming scales (e.g., \citealt{2017ApJ...849...25L,Palau_2018,Beuther_2018,Lu_2020,Morii_2024,Ishihara2024}), whereas others find reduced fragmentation that may be regulated by turbulence (e.g., \citealt{Lu_2015,Rebolledo_2020,Li_2021}) or magnetic fields (e.g., \citealt{Fontani_2016,Beuther_2018b,Saha_2024}). Beyond fragmentation, mass segregation provides a complementary probe of dynamical evolution: it is weak or absent in early-stage IRDCs \citep{Sanhueza_2019_ASHES,Morii2023}, but appears more prominent in evolved protoclusters, where the most massive cores preferentially concentrate toward cluster centers \citep{Dib_2019,Zhang_2022,Xu_2024_ASSEMBLE}. Together with the infall and core-feeding signatures described above, these findings indicate that global gas flows and clump-scale gravity play important roles in shaping dense core populations.

This dynamical picture is consistent with theoretical frameworks such as global hierarchical collapse (GHC; \citealp{Vazquez-Semadeni_2019}) and the inertial-inflow model \citep{Padoan_2020}. Cloud-scale magnetohydrodynamic (MHD) simulations provide further support for this picture. In particular, the STARFORGE simulations follow the formation, accretion, dynamics, and feedback of individual stars within turbulent, magnetized molecular clouds \citep{Grudic2021}. These simulations show that star-forming structures assemble hierarchically while continuing to accrete gas, naturally producing clustered star formation and mass segregation \citep{Guszejnov2022}. These theoretical results provide a framework for interpreting core evolution, but it remains unclear whether the predicted dynamical behavior is common across massive protoclusters.

Chemical diagnostics provide an additional and complementary perspective on dense core evolution by tracing changes in temperature, density, and protostellar activity \citep{van_dishoeck_1998,Jorgensen_2020}. During the early and cold phases, molecular line emission is generally sparse and dominated by simple species. At low temperatures, the freeze-out of CO onto dust grains promotes deuterium fractionation, leading to elevated $\mathrm{N_2D^+/N_2H^+}$ abundance ratios that trace cold, dense gas (e.g., \citealt{Caselli_2002,Crapsi_2005,Emprechtinger_2009,Fontani_2011,Miettinen_2014,Sabatini2022}). As protostellar activity begins, molecular outflows emerge and evolve from highly collimated bipolar structures to wider-angle flows, increasing feedback and dynamical interactions with the surrounding environment (e.g., \citealt{Bachiller_1992, Arce_2006, Bally_2016, Li_2020}). Meanwhile, molecular abundances and excitation conditions evolve significantly as internal heating raises the gas temperature, driving the production of abundant complex organic molecules (COMs) in environments where rapid accretion and rich chemistry coexist (e.g., \citealt{Kurtz_2000,Viti_2004,Herbst_2009_ARAA,Garrod_2022, Qin_2022, Meng2026}). Although chemical evolution serves as a powerful tracer of dense core properties and massive star formation, its variation across evolutionary stages remains insufficiently constrained observationally.

To address these open questions concerning core assembly, fragmentation, and chemical evolution, we carried out the ALMA-QUARKS (Querying Underlying Mechanisms of Massive Star Formation with ALMA-Resolved Gas Kinematics and Structures) survey \citep{Liu_2024_QUARKSI}. As a follow-up to the ATOMS (ALMA Three-millimeter Observations of Massive Star-forming Regions) survey \citep{Liu_2020}, QUARKS observed at 1.3~mm 139 active protoclusters selected from the ATOMS sample, forming one of the largest samples of high-mass star-forming regions observed at high angular resolution. By combining high angular resolution continuum and molecular line observations, QUARKS resolves dense cores down to scales of $\sim10^3$~au at the median distance of the sample and enables statistical investigations of their physical, kinematic, and chemical properties. Compared with other ALMA large programs, QUARKS plus ATOMS contains more targets than ALMA-IMF \citep{Motte2022} and offers multi-scale coverage relative to ALMAGAL \citep{Molinari2025}, thereby providing complementary information to these surveys (e.g., \citealp{Mai_2025}). The QUARKS survey thus offers a unique opportunity to explore how core fragmentation, accretion activity, and chemical differentiation evolve within massive protocluster environments.

This paper is organized as follows. In Sect. \ref{sec:obs}, we describe the observations and data reduction procedures. Sect. \ref{sec:QUARKS_atlas} presents the QUARKS atlas, comprising multi-band and multi-scale images that together provide a complementary view of the ATOMS+QUARKS sample. Sect. \ref{sec:results} presents the identification and classification of dense cores based on molecular line diagnostics and outflow associations. In Sect. \ref{sec:discussion}, we analyze the core mass function and core formation efficiency, and examine the detection rates and chemical properties of different tracers as functions of evolutionary stage and core properties. Finally, Sect. \ref{sec:conclusions} summarizes our main conclusions.


\begin{figure*}[ht!]
\centering
\includegraphics[width=1.0\textwidth]{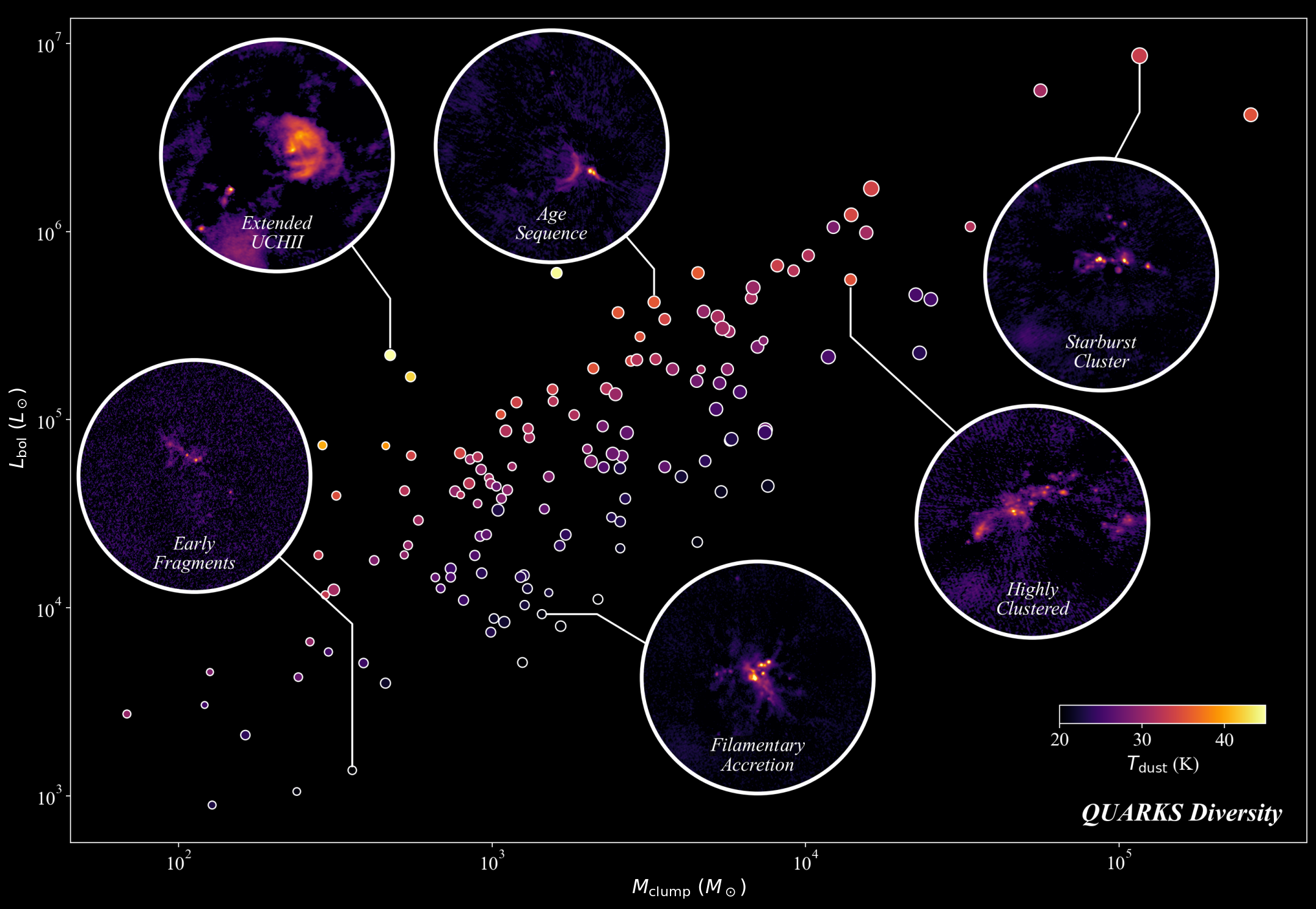} 
\caption{Diversity of the QUARKS sample. The x-axis and y-axis show the clump mass and bolometric luminosity, respectively, and the color of each point represents the clump dust temperature, as indicated by the color bar in the lower-right corner. The sizes of the points represent the clump radii. These clump-scale parameters are adopted from \citet{Liu_2024_QUARKSI}. The dust temperature is used as an approximate evolutionary indicator of the clumps \citep[e.g.,][]{Guzman2015, Giannetti2017}. The white circles highlight representative sources selected from our combined continuum images, illustrating the different fragmentation levels and modes identified within the QUARKS sample.
\label{fig:QUARKS_diversity}}
\end{figure*}

\section{Observations and Data Reduction} \label{sec:obs}
\subsection{ALMA Observations} \label{subsec:obs}

\par As a follow-up to the ATOMS survey, the QUARKS survey (PIs: Lei Zhu, Guido Garay, and Tie Liu; Project ID: 2021.1.00095.S) observed 139 protoclusters from the ATOMS project in Band 6. The detailed sample selection criteria are described in \citet{Liu_2024_QUARKSI}. The luminosity–mass distribution of these protoclusters is shown in Fig.~\ref{fig:QUARKS_diversity}, which indicates their different evolutionary stages. The sample provides an ideal laboratory for obtaining a dynamical view of protocluster evolution.

\par A total of 156 single pointings toward the 139 protoclusters were carried out using three configurations: the ACA 7-m array \citep{Iguchi_2009_ACA}, the ALMA 12 m compact configurations (TM2, C-2/C-3), and the extended configurations (TM1, C-5/C-6), corresponding to synthesized beam sizes of $\sim5^{\prime\prime}$, $\sim1^{\prime\prime}$, and $\sim0.3^{\prime\prime}$, respectively. The QUARKS survey observations are grouped into 15 scheduling blocks (SBs; hereafter ``groups''). All the observations of the QUARKS survey were completed by June 2024. Detailed observational information for the ACA 7-m array and the TM2 observations is provided in \citet{Xu_2024_Quarks} and \citet{Yang_2025b}, respectively. 

\par The TM1 observations were carried out from August 2022 to June 2024, and their observational information is summarized in Table~\ref{tab:TM1_obs} in Appendix~\ref{app:TM1_obs}. Columns 1--3 list the group ID, the number of single-pointing fields, and the sources contained in each group. Each group was observed in 1--3 execution blocks (EBs) on different dates (Column 4) with different numbers of antennas (Column 5). Each pointing was observed for $\sim$150--360 s, with an average integration time of $\sim$250 s (Column 6). Variations in the minimum and maximum baselines, angular resolution (AR), and maximum recoverable scale (MRS) are mainly due to differences in $uv$ coverage, which arise from variations in array configuration, number of antennas, and source elevation (Columns 7--9). The phase and bandpass/flux calibrators used for each group are listed in Columns 10--11.

\par Dual-polarization mode was adopted in the QUARKS TM1 observations. The setup included four wide spectral windows, each with a bandwidth of 1.875 GHz, centered at 217.92, 220.32, 231.37, and 233.52 GHz. Each spectral window consists of 3840 channels, providing a velocity resolution of $\sim$1.3 km s$^{-1}$ at 230 GHz. These four spectral windows cover a variety of molecular lines, including: (1) cold gas tracers (e.g., $\rm{N_2D^+}$); (2) outflow tracers (e.g., CO, SiO, $\rm{H_2CO}$, $\rm{CH_3OH}$); (3) dense gas tracers (e.g., DCN, $\rm{C^{18}O}$, $\rm{HC_3N}$); (4) hot gas tracers (e.g., $\rm{CH_3CN}$, $\rm{NH_2CHO}$); and (5) ionized gas tracers (H30$\alpha$ and He30$\alpha$). More details on the spectral setup and targeted molecular transitions can be found in \cite{Liu_2024_QUARKSI}.

\subsection{Data Reduction and Imaging}

\begin{figure}[t!]
\centering
\includegraphics[width=\columnwidth]{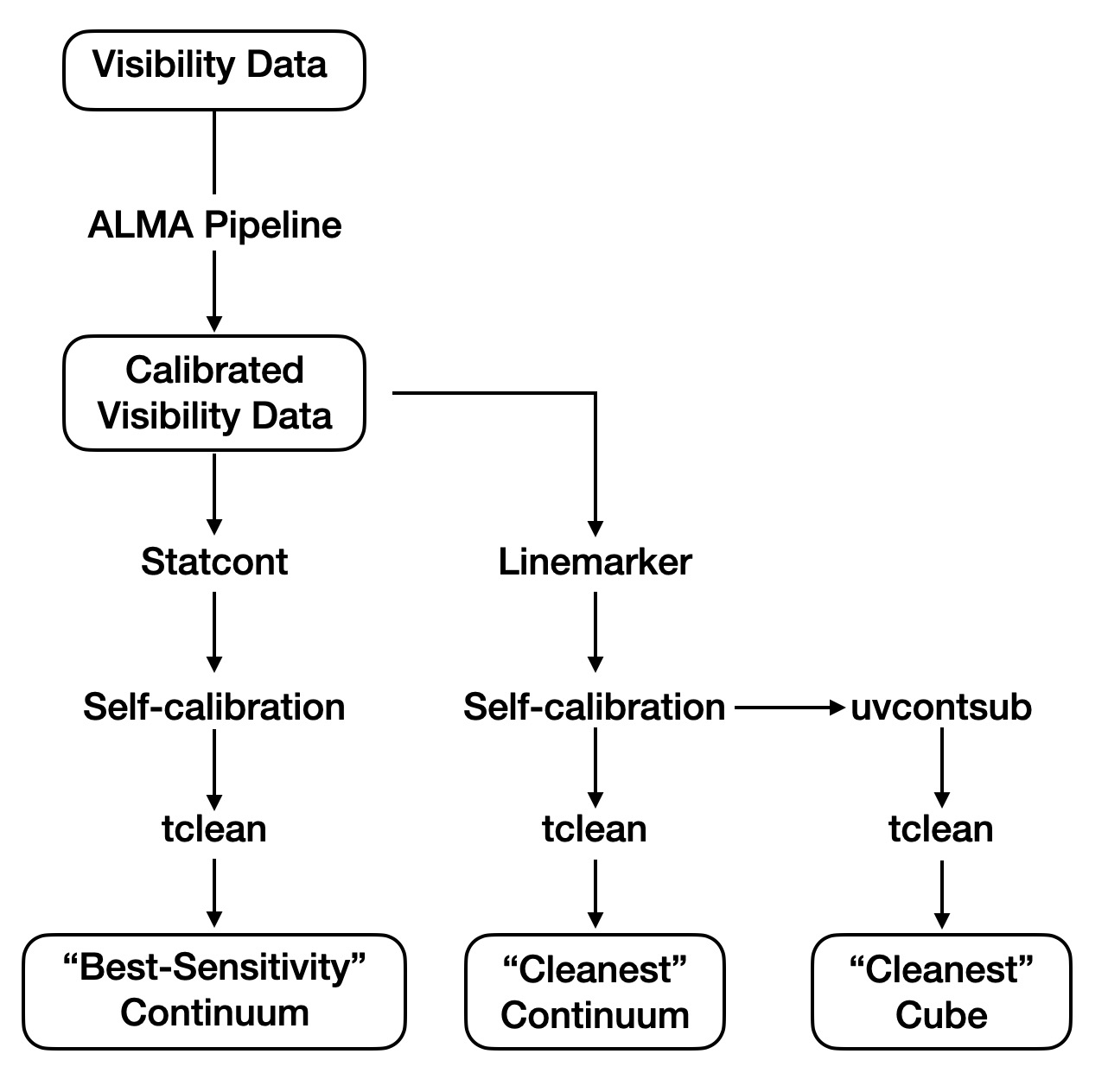}
\caption{Overall workflow chart of the data reduction procedure for QUARKS TM1+TM2+ACA data.}
\label{fig:workflow}
\end{figure}

The visibility data from different configurations were first calibrated using the ALMA pipelines\footnote{https://almascience.nrao.edu/processing/science-pipeline} of the Common Astronomy Software Applications (CASA; \citealt{McMullin_2007_casa1,CASA_Team_2022_casa2}), with versions ranging from 6.2.1 to 6.6.0 depending on the observational dates. The overall data reduction workflow is shown in Fig. \ref{fig:workflow}. The calibrated TM1, TM2, and ACA 7-m visibility data were concatenated into a single measurement set using the \texttt{concat} task.

\par For continuum imaging, following the procedures used in the ALMA-IMF survey \citep{Ginsburg_2022_ALMAIMF_data_reduction}, we produced two sets of continuum products: the ``best sensitivity'' product and the ``cleanest'' product. For the ``best sensitivity'' product, we applied STATCONT \citep{Sanchez-Monge_2018_statcont} to the full-field dirty cubes to identify line-free channels. This approach excluded only channels affected by strong, extended molecular line emission and retained as much bandwidth as possible for continuum imaging. For the ``cleanest'' product, we extracted spectra at the position with the richest molecular line emission and manually identified the line-free channels using the \textit {linemarker}\footnote{https://gitee.com/liuxunchuan/linemarker; a tool for manually identifying line-free channels from spectra} tool. Then, for the ``cleanest'' data, in order to obtain the most reliable image, three rounds of phase-only self-calibration were performed to improve the data quality. We selected the \texttt{hogbom} deconvolution algorithm, which models the sky brightness distribution as a set of point sources and is well suited to iterative phase-only self-calibration. For the ``best sensitivity'' data, we also performed one round of amplitude self-calibration if the signal-to-noise ratio of the continuum image exceeded 200. The continuum imaging process was performed with the \textit{tclean} task, with a Briggs robust weighting of 0.5, resulting in a median synthesized beam size of $0.33^{\prime\prime}\times0.29^{\prime\prime}$. In the cleaning process, the images were automatically masked using the auto-multithresh algorithm, with input parameters recommended by the ALMA guides\footnote{\url{https://casaguides.nrao.edu/index.php/Automasking_Guide_CASA_6.5.4}}. The multi-term multi-frequency synthesis (\texttt{mtmfs}) algorithm was employed for the final \textit{tclean} procedure. We uniformly set the image size to $900 \times 900$ pixels and the cell size to $0.05^{\prime\prime}$ for the single-pointing fields. For the 17 mosaic fields with dual pointings, we changed the image size based on their positions but kept the same cell size. Primary-beam correction was applied with a primary-beam response cutoff of 0.2. We adopted the ``cleanest'' products for subsequent analysis because their more conservative selection of line-free channels minimizes residual line contamination toward chemically rich sources. The ``best sensitivity'' data will be provided upon request.


For spectral lines, we only derived the cubes from the ``cleanest'' visibilities to obtain reliable spectra toward hot cores. We applied all calibration tables derived from the ``cleanest continuum'' imaging process to all four spectral windows. The continuum component was excluded from the visibility data using the \textit{uvcontsub} task. The combined TM1, TM2, and ACA 7-m continuum-subtracted visibility data were imaged with the \textit{tclean} task, using \textit{multiscale} deconvolution to recover possible extended line emission. The resulting spectral-line cubes have a channel width of approximately $0.67~\mathrm{km\,s^{-1}}$. The image size and pixel size were matched to those of the continuum images for an easier comparison in subsequent analyses. Unless otherwise stated, all subsequent analyses in this paper use the combined TM1+TM2+ACA products.

\begin{figure*}
  \centering
  \includegraphics[width=1.0\textwidth]{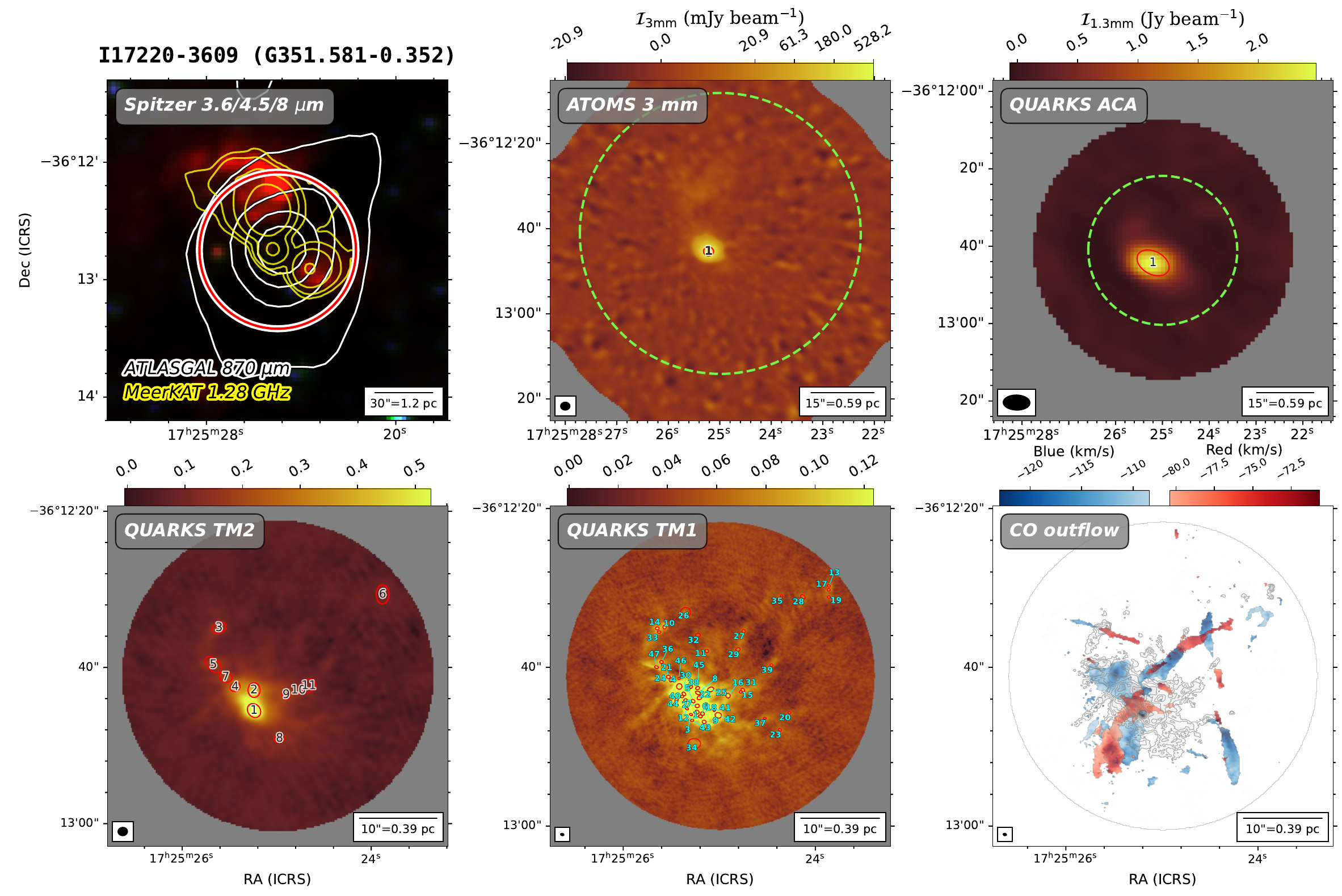}
  \caption{Multi-scale and multi-wavelength atlas of a representative QUARKS source, I17220–3609.  \textbf{Top-left}: The \textit{Spitzer} 3.6/4.5/8~$\mu$m pseudo-color map \citep{Benjamin_2003}, overlaid with ATLASGAL 870~$\mu$m emission \citep{Schuller_2009} as white contours and SARAO MeerKAT Galactic Plane Survey (SMGPS) 1.28~GHz emission \citep{Goedhart_2024} as yellow contours. The red circle marks the adopted ATOMS field of view. \textbf{Top-middle}: ATOMS 3 mm continuum image showing the clump-scale emission. The identified continuum sources are shown as red ellipses \citep{Liu_2021}. The green dashed circle marks the adopted field of view of the QUARKS ACA pointing. \textbf{Top-right}: QUARKS ACA 1.3 mm continuum image. The green dashed circle marks the adopted field of view of the QUARKS 12 m array pointing. The identified continuum sources are marked throughout the atlas with red ellipses ($\rm{S/N}>9$) and yellow ellipses ($\rm{S/N}<9$) \citep{Xu_2024_Quarks}.  \textbf{Bottom-left}: QUARKS TM2+ACA 1.3 mm continuum image. The identified continuum sources are shown as red ellipses \citep{Yang_2025b}. \textbf{Bottom-middle}: QUARKS TM1+TM2+ACA combined 1.3 mm continuum image with the highest angular resolution, resolving substructures within cores. The continuum sources identified in this paper are shown as red ellipses. \textbf{Bottom-right}: QUARKS TM1+TM2+ACA 1.3 mm continuum image shown in grayscale, with logarithmic continuum contours overlaid. The color map shows the velocity field of the CO (2--1) outflow, where blue and red represent the moment-1 maps of the blue- and red-shifted outflow lobes, integrated over velocity ranges of $-150$ to $-105$ km s$^{-1}$ and $-83$ to $-60$ km s$^{-1}$, respectively. For readability, the in-panel labels ``QUARKS TM2'' and ``QUARKS TM1'' are abbreviated forms of the TM2+ACA and TM1+TM2+ACA continuum products, respectively. The color scales in the bottom-left and bottom-middle panels
are in units of Jy~beam$^{-1}$. The synthesized beams are shown in the bottom left corners of the relevant panels.}   
  \label{fig:quarks_atlas}
\end{figure*}

\section{QUARKS Atlas}
\label{sec:QUARKS_atlas}

Following \citet{Xu_2024_Quarks}, we constructed a comprehensive ATOMS+QUARKS atlas that provides a unified multiscale and multiphase view of high-mass star formation. By combining infrared, radio, and multi-configuration ALMA observations, the ATOMS+QUARKS data set bridges spatial scales from the parsec-scale clump environment down to $\sim10^3$ au dense cores, offering a coherent picture of the hierarchical assembly of massive protoclusters.

On large scales, Spitzer infrared images trace the embedded stellar population, warm dust, and photodissociation-region emission, while MeerKAT 1.28 GHz radio images primarily trace ionized gas and the overall star-forming environment of each clump. Submillimeter continuum data from the ATLASGAL survey reveal the distribution of cold, dense gas that defines the parent clump structure. At intermediate scales, the ATOMS 3\,mm continuum maps identify dense clump substructures and hub-filament systems at $\sim0.1$\,pc scales, providing a direct link between the clump-scale mass reservoir and the sites of active star formation. 

The QUARKS 1.3\,mm observations further resolve the internal substructures within these regions. The ACA data trace subclump-scale fragments, while the higher-resolution TM2 and TM1 configurations gradually resolve dense cores and compact condensations embedded within protoclusters. Additionally, representative molecular-line maps, such as CO maps revealing molecular outflows, highlight the associated dynamical activities and feedback processes driven by embedded protostars.

By combining tracers of different physical components, including dense molecular gas, cold dust, ionized gas, and molecular outflows, the ATOMS+QUARKS atlas reveals the complex structures, dynamics, and chemistry of massive star-forming regions. These multi-scale images not only visualize how dense gas is concentrated and fragmented at different physical scales, but also provide environmental information for interpreting the properties of dense cores discussed in the following sections. An example of the multi-scale and multi-wavelength atlas is shown in Fig.~\ref{fig:quarks_atlas}. The complete atlas and continuum images are available in the online material\footnote{
\href{https://almaquarks.github.io/}{https://almaquarks.github.io/}.}.

\section{Results} \label{sec:results}

\subsection{Source Extraction Algorithm}

Several algorithms are commonly used to extract compact sources from continuum images, including \texttt{CLUMPFIND} (\citealp{Williams_1994_clumpfind}; e.g., \citealp{Enoch_2006}; \citealp{Di_francesco_2008}), \texttt{astrodendro} (\citealp{Rosolowsky_2008_astrodendro}; e.g., \citealp{Sanhueza_2019_ASHES}; \citealp{Coletta_2025_ALMAGAL}), \texttt{CuTEx} (\citealp{Molinari_2011_cutex}; e.g., \citealp{Molinari_2016}; \citealp{Elia_2017_Higal}), and \texttt{getsf} (\citealp{Menshchikov_2021_getsf}; e.g., \citealp{Xu_2024_ASSEMBLE}; \citealp{Louvet_2024_ALMAIMF}). These algorithms adopt different extraction strategies and have different strengths and limitations, as summarized in Appendix~\ref{app:source_extraction}.

\par In this work, we used \texttt{getsf} to extract the dense cores for the following reasons: 1. As shown in Fig.~\ref{fig:QUARKS_diversity}, the physical properties and environments vary significantly among the QUARKS sources. As summarized in Appendix~\ref{app:source_extraction}, \texttt{getsf} separates compact sources from filamentary and diffuse emission, making it well suited to the complex environments in our sample. 2. The distances of the sources in the QUARKS survey span 1.2--12.9~kpc \citep{Liu_2024_QUARKSI}, resulting in large variations in spatial resolution. \texttt{getsf} maintains good consistency across images with different resolutions. 3. \texttt{getsf} is sensitive to sources over a range of spatial scales and can detect both compact and diffuse sources. 4. In an earlier paper of the QUARKS series, \citet{Yang_2025b} used \texttt{getsf} to extract sources from the TM2+ACA continuum images. To ensure consistency in subsequent studies (e.g., hierarchical fragmentation), we therefore adopted the same source-extraction algorithm in this work. To accelerate the process, we employed GNU parallel \citep{Tange_2024}.

\par We ran \texttt{getsf} on the continuum images without primary-beam correction to maintain a uniform noise level across each image, and subsequently measured the flux densities after applying the primary-beam correction. In the configuration file, we set the maximum source size to five times the synthesized beam size, which is slightly larger than the TM2 angular resolution. To identify as many barely resolved sources as possible, we adopted a minimum source size of 0.9 times the beam size. To ensure the robustness of the source extraction, we retained only those cores with a significance level (\textit{SIGNM}; detection significance from monochromatic single scales) greater than 5 and a goodness (\textit{GOODM}; monochromatic goodness combining significance and signal-to-noise ratio) larger than 1, consistent with the parameters reported in \citet{Yang_2025b}. We then visually inspected the sources located in regions where the primary-beam response was below 0.3 and excluded several weak detections at the edge of the image. Finally, we excluded the sources whose peak or total flux was negative. In total, we identified 3,374 compact sources, whose basic physical properties are summarized in Table~\ref{tab:sources}.

\begin{deluxetable*}{lccccccccccc}
\tablecaption{Basic Measurements of TM1 Compact Continuum Sources\label{tab:sources}}
\tablewidth{0pt}
\tablehead{
\colhead{Source} & \colhead{ID} & \colhead{Distance} &
\colhead{R.A. (ICRS)} & \colhead{Decl. (ICRS)} &
\colhead{$\theta_{\rm maj}$} & \colhead{$\theta_{\rm min}$} & \colhead{PA} &
\colhead{$I_{\rm peak}$} & \colhead{$F_{\rm int}$} & \colhead{SIGNM} & \colhead{GOODM} \\
\colhead{} & \colhead{} & \colhead{(kpc)} &
\colhead{(hh:mm:ss.sss)} & \colhead{(dd:mm:ss.ss)} &
\colhead{(arcsec)} & \colhead{(arcsec)} & \colhead{(deg)} &
\colhead{(mJy beam$^{-1}$)} & \colhead{(mJy)} & \colhead{} & \colhead{} \\
\colhead{(1)} & \colhead{(2)} & \colhead{(3)} & \colhead{(4)} &
\colhead{(5)} & \colhead{(6)} & \colhead{(7)} & \colhead{(8)} &
\colhead{(9)} & \colhead{(10)} & \colhead{(11)} & \colhead{(12)}}
\startdata
I08303-4303 & 1 & 2.35 & 08:32:08.643 & -43:13:45.86 & 0.674 & 0.528 & 166.0 & 19.29(0.47) & 69.63(0.79) & 181.0 & 860.9 \\
I08303-4303 & 2 & 2.35 & 08:32:09.039 & -43:13:43.40 & 0.379 & 0.336 & 95.3 & 12.94(0.63) & 16.07(0.48) & 121.8 & 282.6 \\
I08303-4303 & 3 & 2.35 & 08:32:08.659 & -43:13:45.32 & 0.462 & 0.389 & 93.6 & 12.45(0.47) & 23.79(0.55) & 71.0 & 202.8 \\
I08303-4303 & 4 & 2.35 & 08:32:08.460 & -43:13:47.28 & 0.328 & 0.325 & 84.9 & 12.2(1.0) & 12.03(0.70) & 81.9 & 116.4 \\
I08303-4303 & 5 & 2.35 & 08:32:08.456 & -43:13:49.19 & 0.451 & 0.364 & 62.1 & 7.96(1.00) & 11.68(0.82) & 35.0 & 30.1 \\
I08303-4303 & 6 & 2.35 & 08:32:08.791 & -43:13:46.28 & 0.313 & 0.287 & 42.2 & 2.98(0.28) & 2.38(0.17) & 24.9 & 28.4 \\
I08303-4303 & 7 & 2.35 & 08:32:08.891 & -43:13:52.27 & 0.568 & 0.492 & 35.9 & 1.67(0.21) & 4.35(0.24) & 13.4 & 13.8 \\
I08303-4303 & 8 & 2.35 & 08:32:09.757 & -43:13:48.47 & 0.358 & 0.308 & 44.6 & 1.20(0.18) & 1.23(0.13) & 9.6 & 6.6 \\
I08303-4303 & 9 & 2.35 & 08:32:10.159 & -43:13:46.03 & 2.006 & 1.663 & 91.2 & 0.43(0.18) & 8.16(0.48) & 10.8 & 5.7 \\
I08303-4303 & 10 & 2.35 & 08:32:08.687 & -43:13:44.50 & 0.728 & 0.651 & 3.9 & 1.76(0.51) & 5.91(0.45) & 8.3 & 5.0 \\
\enddata
\tablecomments{Source names and continuum source IDs are listed in Columns
(1)--(2). The source kinematic distances are given in Column (3). The ICRS coordinates are listed in Columns (4)--(5). The convolved FWHM values of the major and
minor axes, and the position angle (PA), are listed in Columns (6)--(8). The peak intensity and integrated flux, with uncertainties in parentheses, based on the measurements derived by \texttt{getsf}, are given in Columns (9)--(10). The detection significance (SIGNM) and the monochromatic quality indicator (GOODM), as derived by \texttt{getsf}, are listed in Columns (11)--(12). This table shows the first 10 entries as an example. The table is available in its entirety in machine-readable form.}
\end{deluxetable*}

\subsection{Velocity Identification}
\label{sec:core_velocity}

To determine the systemic velocity of each compact core, we primarily relied on Gaussian fitting of the C$^{18}$O (2--1) line profiles because this transition has a high detection rate in our sample and generally traces molecular gas kinematically associated with dense cores \citep{Kirk_2007}. For each core, we extracted the core-averaged spectrum centered on the C$^{18}$O frequency and searched for emission peaks within a velocity window of $\pm 20~\mathrm{km ~s^{-1}}$ around the clump systemic velocity. If a positive emission peak exceeded the $3.5\sigma$ threshold, we performed a Gaussian fit to the spectrum, and a detection was considered reliable only when the fitted amplitude was greater than $3.5\sigma$.

A substantial fraction of the cores exhibit multiple C$^{18}$O velocity components. Closely spaced components were fitted with a single Gaussian, yielding a centroid between them. This treatment does not significantly affect the line detection rates because the fitted velocity was used primarily to define the search window for other molecular lines, whose velocity components generally coincide with those of C$^{18}$O. For well-separated components or components with substantially different intensities, we visually inspected the spectra and selected the dominant component consistent with the other molecular emission lines detected toward the core.

Because cores near H{\sc ii} regions may exhibit absorption features in their spectra, we simultaneously searched for absorption peaks below $-5\sigma$. In such cases, a negative Gaussian was fitted to capture the absorption profile. If only an emission feature or only an absorption feature was present, the centroid velocity of the corresponding fitted Gaussian was adopted as the systemic velocity of the core. A small fraction of sources exhibited both emission and absorption signatures (520 cores, 15.4\% of the full sample). These spectra were visually inspected, and the systemic velocity was determined based on the velocities of other molecular lines detected toward the same core.

For cores that show no significant C$^{18}$O emission or absorption, we performed Gaussian fits to the H$_2$CO $3_{0,3}-2_{0,2}$ ($E_u/k=21.0$ K) or the CH$_3$CN $J_K=12_4-11_4$ ($E_u/k = 183.1~\mathrm{K}$) emission lines. If either the H$_2$CO or the CH$_3$CN line provided a reliable detection, its fitted centroid velocity was adopted. Otherwise, the systemic velocity was recorded as ``NaN'', and the clump systemic velocity adopted from \citet{Liu_2024_QUARKSI} was used for that core in subsequent analyses. The derived velocity information is summarized in Table~\ref{tab:tm1_table3}.

\subsection{Evolutionary Classification of Cores}

\par Core classification provides crucial insights into the evolutionary sequence of dense cores. As discussed in Sect. \ref{subsec:obs}, the QUARKS survey includes four wide spectral windows with abundant molecular line tracers spanning a range of excitation temperatures, which help distinguish cores at different evolutionary stages. This classification enables more accurate estimates of physical properties such as mass and density and reveals how these properties evolve over time.

\par
\begin{figure}[t!]
\centering
\includegraphics[width=\columnwidth]{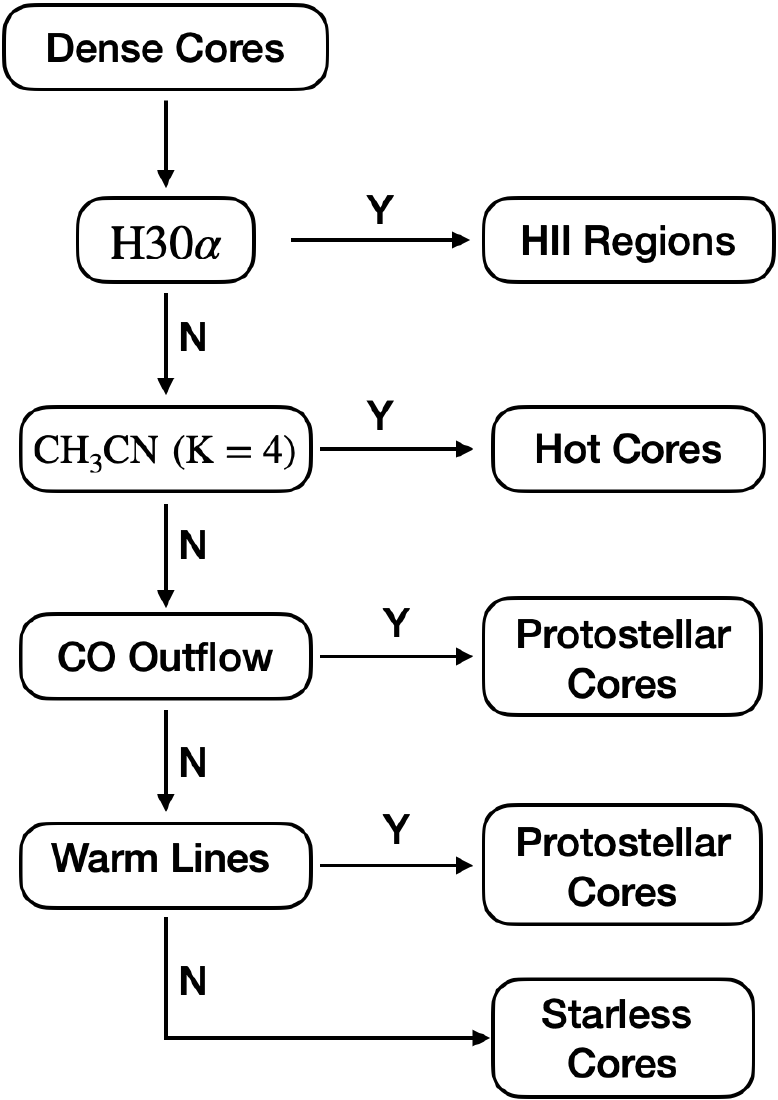}
\caption{The overall criteria for core classification. In this work, the identified dense cores are classified into four categories: starless core candidates, protostellar core candidates, hot core candidates, and H{\sc ii} region candidates.}
\label{fig:core_classification}
\end{figure}

 In an earlier QUARKS study, \citet{Yang_2025b} classified the TM2+ACA cores mainly based on the number of detectable molecular lines ($N_{\rm line}$). In this work, we adopt a different approach by selecting specific molecular lines with different upper-level energies and using their detections to classify the cores into different evolutionary categories. Because the cores in our sample are at different evolutionary stages and embedded in different environments, some of them exhibit highly complex spectral characteristics. Therefore, all classified cores should be regarded as \textbf{candidates}. We classified the identified cores into four categories: starless core candidates, protostellar core candidates, hot core candidates, and H{\sc ii} region candidates. The classification criteria are illustrated in Fig.~\ref{fig:core_classification}. For simplicity, the term ``candidates'' will be omitted hereafter.

\subsubsection{H{\sc ii} Region Identification} \label{subsubsec:HII_identification}
H{\sc ii} regions are typically identified through ionized-gas tracers, including radio recombination lines such as H30$\alpha$ and free-free continuum emission (e.g., \citealt{Walsh_1998, Yang_2019, Yang_2021, Galvan-madride_2024_ALMAIMF}). Since the QUARKS survey does not include dedicated free-free continuum observations, we identified H{\sc ii} regions based on the detection of the H30$\alpha$ line. Following the method described in \citet{Nony_2024}, we extracted core-averaged spectra around the $\rm{H30\alpha}$ frequency and searched for emission peaks within a velocity range of $\pm30$ km s$^{-1}$ around the clump systemic velocity. The choice of 30 km s$^{-1}$ was motivated by the fact that there may be a velocity difference of more than 20 km s$^{-1}$ between the $\rm{H30\alpha}$ line and the clump systemic velocity \citep{Zhang_2026}. If a peak exceeded the $3.5\sigma$ threshold, we performed a Gaussian fit to the spectrum. As radio recombination lines are expected to be much broader than molecular lines \citep{Hoare_2007}, we required the fitted $\rm{FWHM}$ to be larger than 20 km s$^{-1}$ and the fitted amplitude to exceed $3.5\sigma$ to minimize contamination by molecular lines. Based on these criteria, we identified 379 continuum sources associated with H30$\alpha$ emission, which we classify as H{\sc ii} region candidates. Because multiple 1.3~mm continuum peaks within a single extended H{\sc ii} region may be classified separately, the number of physically distinct H{\sc ii} regions requires further verification using the spatial distribution and moment maps of the H30$\alpha$ line emission in future work.

\subsubsection{Hot Core Identification}
\label{subsubsec:hotcore_identification}
Hot cores are objects in the early stages of high-mass star formation, characterized by high temperatures (${T}>100~\rm{K}$) and complex molecular compositions \citep{Kurtz_2000, Herbst_2009_ARAA}. Studying hot cores is essential for understanding the chemical enrichment of the interstellar medium (e.g., \citealp{Charnley_1992, Garrod_2006}) and provides valuable insights into the early stages of massive star formation \citep{Cesaroni_2005}. However, there is no universally agreed-upon definition of hot cores, and previous studies have identified them using different gas tracers, such as $\rm{CH_3OCHO}$ \citep{Bonfand_2024} and $\rm{C_2H_5CN}$ \citep{Qin_2022}. 

In this work, we identified hot cores using the CH$_3$CN $J=12_4-11_4$ ($E_u/k = 183.1~\mathrm{K}$) line, a commonly used high-excitation tracer of warm, dense gas \citep[e.g.,][]{Hernandez_2014,Taniguchi_2024}. We applied selection criteria analogous to those used for H{\sc ii} regions, except that no linewidth criterion was imposed because molecular lines such as CH$_3$CN are generally narrower than radio recombination lines \citep[e.g.,][]{Purcell2006}.  To ensure reliable detections and avoid confusion with unrelated molecular features, we required (1) both the peak intensity and fitted amplitude to exceed $3.5\sigma$,  and (2) a fitted centroid velocity within $\pm 5~\mathrm{km\,s^{-1}}$ of the systemic velocity of the core derived in Sect.~\ref{sec:core_velocity}. Based on these criteria, we identified 470 hot cores ($13.9\%$).

\par However, the criteria adopted here are relatively simple. A more robust hot-core sample could be constructed by examining the spatial distribution of complex organic molecule emission in moment-0 maps and deriving rotational temperatures by fitting multiple molecular transitions (e.g., \citealt{Qin_2022}). These analyses would help exclude spurious detections introduced by source-extraction algorithms and ensure that only genuine hot cores are included. These enhanced selection methods will be explored in future work (S.-L. Qin et al. in prep.).

\begin{figure}[t!]
\centering
\includegraphics[width=\columnwidth]{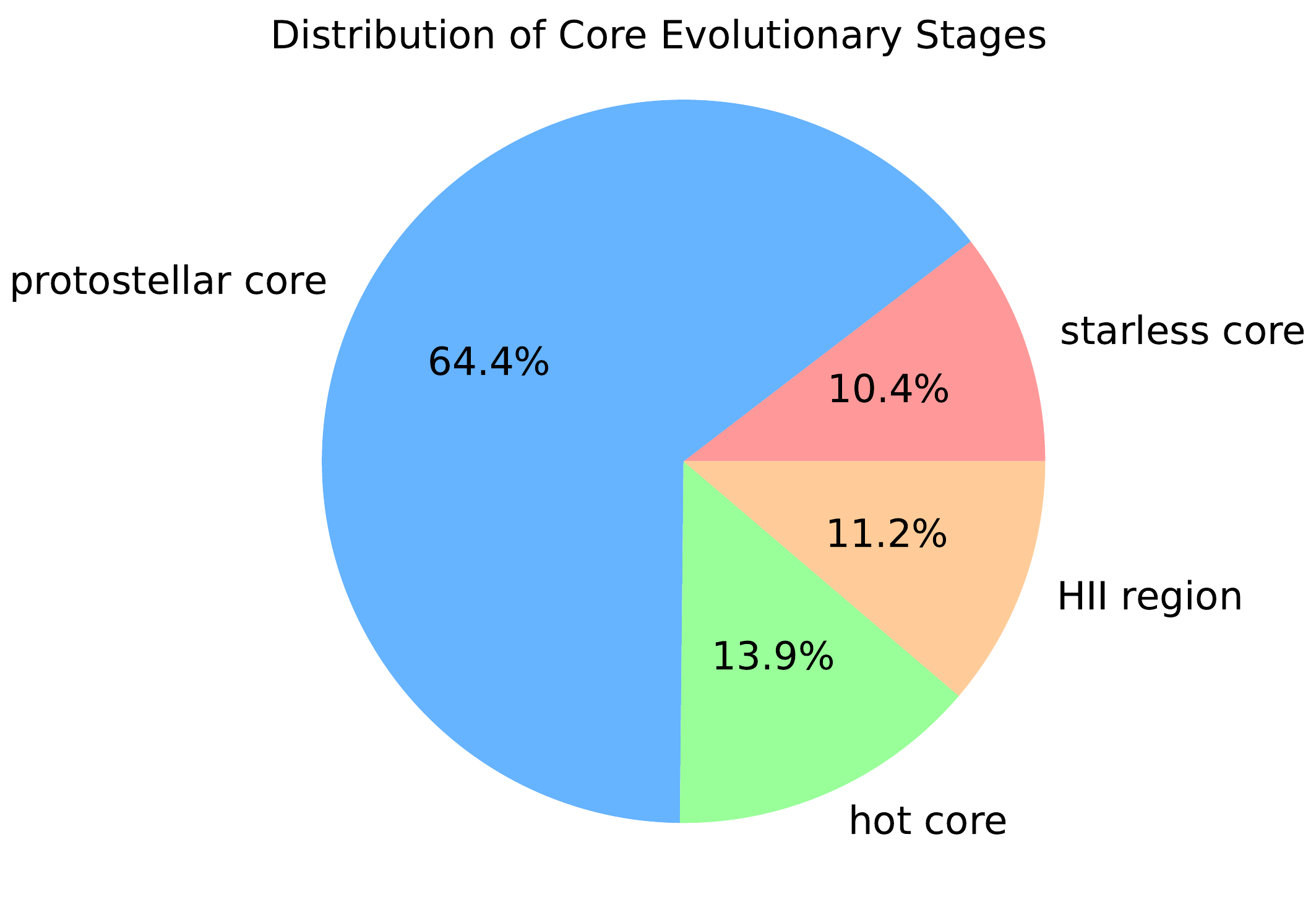}
\caption{Overall distribution of the 3374 cores across the four evolutionary categories. The sample contains 351 starless core candidates (10.4\%), 2174 protostellar core candidates (64.4\%), 470 hot core candidates (13.9\%), and 379 H{\sc ii} region candidates (11.2\%).}
\label{fig:core_fraction}
\end{figure}

\subsubsection{CO Outflow Association}
\label{subsubsec:CO_outflow}

\par The presence of outflows is one of the first manifestations of star formation activity \citep{Bachiller_1996}. Outflows provide crucial information about mass accretion, with more massive stars typically driving more powerful outflows (e.g., \citealp{Beuther_2002, Zhang_2005, Yang_2018}). Investigating whether a core is associated with outflow activity helps distinguish between starless and protostellar stages (e.g., \citealp{Feng_2016, Svoboda_2019, Li_2020}). In the QUARKS survey, we have two widely used outflow tracers: CO (2-1) and SiO (5-4). Since CO is more abundant and sensitive to low-density gas, it is effective for detecting weak and spatially extended molecular outflows. We therefore use the presence of associated CO outflow emission to identify protostellar cores.

CO outflows often exhibit a wide variety of morphologies, ranging from collimated jets to wide-angle winds (e.g., \citealp{Arce_2007, Bally_2016}). As a result, it can be difficult to determine whether a dense core is associated with a CO outflow. In this work, we visually inspected the CO channel maps around each core relative to the core systemic velocity derived in Sect.~\ref{sec:core_velocity} to assess the outflow association. We classified the outflow associations into four categories: Bipolar (B), Unipolar (U), Confused (C), and No association (N). In the QUARKS survey, most of the cores are located in clustered environments, making it difficult to pinpoint the driving source of the outflows. Moreover, a large fraction of the detected outflows are wide-angle outflows, which can produce complex CO emission morphologies. To address these challenges, we adopt two basic rules: 1. When the driving source is ambiguous, we provisionally associate the outflow with the brightest nearby continuum source. 2. We identify a single outflow lobe by requiring continuous CO emission across more than 5 channels, corresponding to a velocity range of at least $\sim3.3~\mathrm{km\,s^{-1}}$. For cores embedded in strong, spatially extended CO outflow emission,
or surrounded by CO lobes whose velocity offsets from the core systemic velocity satisfy $|v_{\rm diff}| <5~\mathrm{km\,s^{-1}}$, we mark the core as ``C'' (Confused). Cores classified as ``B'' (Bipolar) or ``U'' (Unipolar) are characterized as protostellar cores. In total, we identified 257 cores with bipolar outflows, 193 sources with unipolar outflows, 1454 sources with confused outflow associations, and 1470 sources with no outflow association. Detailed studies on outflows in the whole sample will be presented in a forthcoming paper (Y. F. Jiao et al. in preparation).

\subsubsection{Protostellar and Starless Core Identification}

As CO emission is often spatially confused in clustered regions, we also used the detection of specific molecular tracers with relatively high upper-level energies to distinguish between starless and protostellar cores. CO outflow associations and warm-line detections are not treated as separate evolutionary categories. Instead, they provide complementary indicators of protostellar activity. Following \citet{Sanhueza_2019_ASHES}, we adopted typical warm molecular transitions with upper-level energies exceeding 22 K, including ${\rm H_2CO}~J=3_{2,2}-2_{2,1}$ ($E_u/k=68.09~\rm K$), ${\rm H_2CO}~J=3_{2,1}-2_{2,0}$ ($E_u/k=68.11~\rm K$), and ${\rm CH_3OH}~J_k=4_{2,2}-3_{1,2}$ ($E_u/k=45.46~\rm K$). In the QUARKS survey, we added one additional warm transition, ${\rm SO}~J=6_5-5_4$ ($E_u/k=34.98~\rm K$), which is widely used as a tracer of protostellar activity, including infall, shocks, and outflows \citep{Lee_2006,Silva_2017}. The same detection method described in Sect.~\ref{subsubsec:hotcore_identification} was applied, except that a wider centroid-velocity tolerance of $\pm10$~km~s$^{-1}$ was adopted for SO because its line profiles are typically broader. Cores associated with any of these warm lines were classified as protostellar cores. Note that some starless cores located near hot cores may also show these lines due to external heating. To minimize contamination of the starless-core sample by protostellar sources, we adopted this conservative criterion, which may overestimate the number of protostellar cores. After checking both outflow associations and warm-line detections, we identified 2174 protostellar cores ($64.4\%$). The remaining 351 cores, which show neither clear CO outflow associations nor detections of the selected warm molecular tracers, were classified as starless cores ($10.4\%$). The fractions of cores at different evolutionary stages are shown in Fig.~\ref{fig:core_fraction}.


\begingroup
\setlength{\tabcolsep}{14pt}
\begin{deluxetable*}{@{\extracolsep{\fill}}lcccccccc}
\tablecaption{Physical Properties of TM1 Compact Continuum Cores\label{tab:tm1_table3}}
\tablewidth{0pt}
\tablehead{
\colhead{Source} &
\colhead{ID} &
\colhead{Outflow} &
\colhead{Velocity} &
\colhead{Class} &
\colhead{$T_{\mathrm{d}}$} &
\colhead{$M_{\mathrm{core}}$} &
\colhead{$R_{\mathrm{dec}}$} &
\colhead{$n(\mathrm{H}_2)$} \\
\colhead{(1)} &
\colhead{(2)} &
\colhead{(3)} &
\colhead{(4)} &
\colhead{(5)} &
\colhead{(6)} &
\colhead{(7)} &
\colhead{(8)} &
\colhead{(9)} \\
\colhead{} &
\colhead{} &
\colhead{} &
\colhead{(km\,s$^{-1}$)} &
\colhead{} &
\colhead{(K)} &
\colhead{($M_\odot$)} &
\colhead{(au)} &
\colhead{($10^6$ cm$^{-3}$)}
}
\startdata
I08303-4303 & 1 & B & 15.5$^{a}$ & 3 & 100 & 1.4(0.6) & 1170 & 26(13) \\
I08303-4303 & 2 & B & 13.3$^{a}$ & 2 & 30 & 1.2(0.5) & 390 & 630(310) \\
I08303-4303 & 3 & C & 12.2$^{a}$ & 3 & 100 & 0.5(0.2) & 630 & 59(29) \\
I08303-4303 & 4 & C & 14.3$^{a}$ & 2 & 30 & 0.9(0.4) & 390 & 470(230) \\
I08303-4303 & 5 & B & 15.4$^{a}$ & 3 & 100 & 0.2(0.1) & 550 & 42(21) \\
I08303-4303 & 6 & C & 13.5$^{a}$ & 2 & 30 & 0.18(0.08) & 390 & 94(46) \\
I08303-4303 & 7 & C & 14.1$^{a}$ & 2 & 30 & 0.3(0.1) & 970 & 10.9(5.4) \\
I08303-4303 & 8 & N & 14.3$^{e}$ & 1 & 20 & 0.15(0.07) & 390 & 80(40) \\
I08303-4303 & 9 & N & 12.1$^{a}$ & 2 & 30 & 0.6(0.3) & 4220 & 0.2(0.1) \\
I08303-4303 & 10 & C & 12.2$^{a}$ & 3 & 100 & 0.12(0.05) & 1420 & 1.2(0.6) \\
\enddata
\tablenotetext{a}{Velocity from $\mathrm{C^{18}O}$ emission line without absorption.}
\tablenotetext{b}{Velocity from $\mathrm{C^{18}O}$ emission line with absorption.}
\tablenotetext{c}{Velocity from $\mathrm{C^{18}O}$ absorption line.}
\tablenotetext{d}{Velocity from $\mathrm{H_2CO}$ or $\mathrm{CH_3CN}$ emission lines.}
\tablenotetext{e}{No reliable core-scale spectral line; the systemic velocity of the clump is adopted.}
\tablecomments{Columns: (1) Source name; (2) core ID; (3) CO outflow association (B = bipolar, U = unipolar, C = confused, N = no association); only cores with B or U are considered to be associated with outflow activity; (4) core velocity (see table notes); (5) evolutionary classification: 1 = starless core, 2 = protostellar core, 3 = hot core, 4 = H{\sc ii} region; (6) assumed dust temperature assigned according to the evolutionary classification; (7) core mass derived from the 1.3\,mm continuum emission; (8) deconvolved equivalent radius; and (9) H$_2$ number density in units of $10^6$\,cm$^{-3}$. Values in parentheses in Columns (7) and (9) are the estimated $1\sigma$ uncertainties in the same units as the corresponding quantities. The quoted uncertainties do not include the uncertainty in the adopted dust temperatures. This table shows the first 10 entries as an example, and the full table is available in machine-readable form.}
\end{deluxetable*}
\endgroup

\begin{figure*}[htbp]
    \centering
    \includegraphics[width=\textwidth]{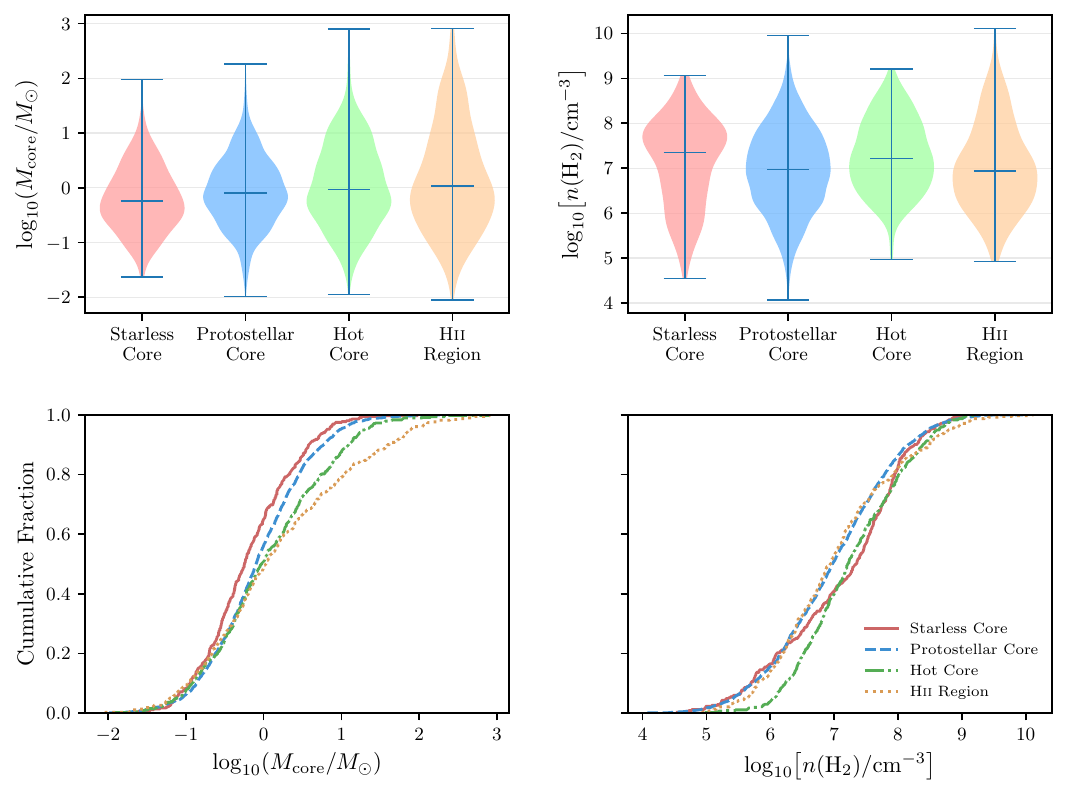}
    \caption{Violin plots (top row) and empirical cumulative distribution functions (bottom row) showing the distributions of core mass (left column) and molecular hydrogen number density (right column) across different evolutionary stages. Top: The width of each violin represents a kernel density estimate of the distribution. The horizontal bars indicate the median values, while the vertical lines and end caps span the minimum-to-maximum ranges. Bottom: The empirical cumulative distribution functions show the cumulative fraction of cores below a given value. The colors and line styles denote the four evolutionary stages, as listed in the legend.}
    \label{fig:physical_parameter_evolution}
\end{figure*}

\subsection{Physical Property Estimation}
\label{sec:physical_property}

After classifying the cores, we can estimate their physical properties. We adopt dust temperatures of 20 K for starless cores, 30 K for protostellar cores, and 100 K for hot cores and H{\sc ii} regions \citep[e.g.,][]{Motte2022, Louvet_2024_ALMAIMF, Yang_2025b}. Note that the temperature of each core can be more accurately estimated through rotational temperature fitting when suitable molecular line emission is detected \citep[e.g.,][]{Battersby_2014}. In this work, however, we simply assign the same dust temperature to all cores within each evolutionary stage. Assuming optically thin dust emission, the gas mass of the cores can be calculated using the following equation:
\begin{equation}
M_{\rm core} = R \frac{F_{\rm int} D^{2}}{\kappa_{\nu} B_{\nu}(T)},
\end{equation}
where $F_{\rm int}$ is the integrated flux of the source, $R = 100$ is the gas-to-dust mass ratio, $D$ is the distance to the source, $\kappa_{\nu}$ is the dust opacity, and $B_{\nu}$ is the Planck function at the dust temperature $T$. We adopt a dust opacity of $\kappa_{1.3\,\rm mm} = 0.9~\rm cm^{2}~g^{-1}$, which corresponds to dust grains with thin ice mantles at a gas density of $10^{6}~\rm cm^{-3}$ \citep{OH_1994}. Note that the continuum emission of H{\sc ii} regions is strongly affected by free–free emission, which can lead to an overestimation of their masses \citep[e.g.,][]{Nony_2024}. The correction for free–free emission will be investigated further using multi-wavelength data in future work.

\par The deconvolved size of each core is calculated as
$R_{\rm{dec}}=\sqrt{\theta_{\rm{maj}} \theta_{\rm{min}}-\theta_{\rm{beam}}^{2}} \times D$ \citep{Pouteau_2022}. To avoid deconvolution artifacts, we set the minimum angular size of each core to be $\frac{1}{2}\theta_{\rm{beam}}$. The molecular hydrogen number density is then calculated as 
$n({\mathrm{H}_2}) = \frac{M_{\mathrm{core}}}{\frac{4}{3} \pi \mu m_{\mathrm{H}} R_{\mathrm{dec}}^{3}}$, 
where $\mu = 2.8$ is the molecular weight per hydrogen molecule \citep{Kauffmann_2008} and $m_{\rm H}$ is the mass of a hydrogen atom.

We estimated the uncertainties in $M_{\rm core}$ and $n({\rm H}_2)$ through standard error propagation. Following the uncertainty
estimation procedure of \citet{Sanhueza_2019_ASHES}, we adopted fractional uncertainties of 10\% for both the distance and absolute flux calibration, 23\% for the gas-to-dust mass ratio, and 28\% for the dust opacity. We also assumed a representative 10\% uncertainty
in the deconvolved angular scale and included the individual integrated-flux uncertainty reported by \texttt{getsf}. The resulting median fractional uncertainties are approximately 43\% and 50\% for
$M_{\rm core}$ and $n({\rm H}_2)$, respectively. 

\par Note that these uncertainty estimates do not include the uncertainty in the adopted dust temperatures. As an illustrative estimate, varying the adopted temperatures by $\pm20\%$ changes the
derived masses and number densities by approximately 17--35\% at 225 GHz over the adopted temperature range of 20--100 K, with lower temperatures producing larger inferred values. Future work will derive more accurate core temperatures through multi-transition molecular-line fitting and use them to refine the derived physical properties.

The derived physical properties of the cores are listed in Table~\ref{tab:tm1_table3}. Figure~\ref{fig:physical_parameter_evolution}
shows the distributions of $M_{\rm core}$ and $n({\rm H}_2)$ across evolutionary stages. We used the Jonckheere--Terpstra test to assess monotonic trends along the evolutionary sequence \citep{Terpstra_1952, Jonckheere_1954}. Under the adopted temperature assumptions, $M_{\rm core}$ shows an 
increasing trend
($p=4.4\times10^{-10}$), whereas $n({\rm H}_2)$ does not show a clear trend across the evolutionary stages ($p=0.166$).

\section{Discussion} \label{sec:discussion}

\subsection{Core Mass Function in Protoclusters}

The initial mass function (IMF) describes the distribution of stellar masses at birth over a range of approximately $0.08$--$100\,M_\odot$ and plays an important role in star formation and galaxy evolution. Observationally, the IMF exhibits a log-normal-like shape at low masses and a power-law tail at the high-mass end. The high-mass IMF is commonly described by the Salpeter power-law slope, $dN/d\log M \propto M^{-1.35}$ \citep{Salpeter_1955}. This slope provides a widely used empirical reference and is broadly consistent with many measurements \citep{Bastian_2010,Kroupa_2013}; we therefore adopt it as the canonical reference in this work. Nevertheless, the universality of the IMF shape remains under debate \citep[e.g.,][]{Dib_2014,Dib_2017,Schneider_2018,Mor_2019, Dib_2023}.

\begin{figure*}[t]
    \centering
    \includegraphics[width=\textwidth]{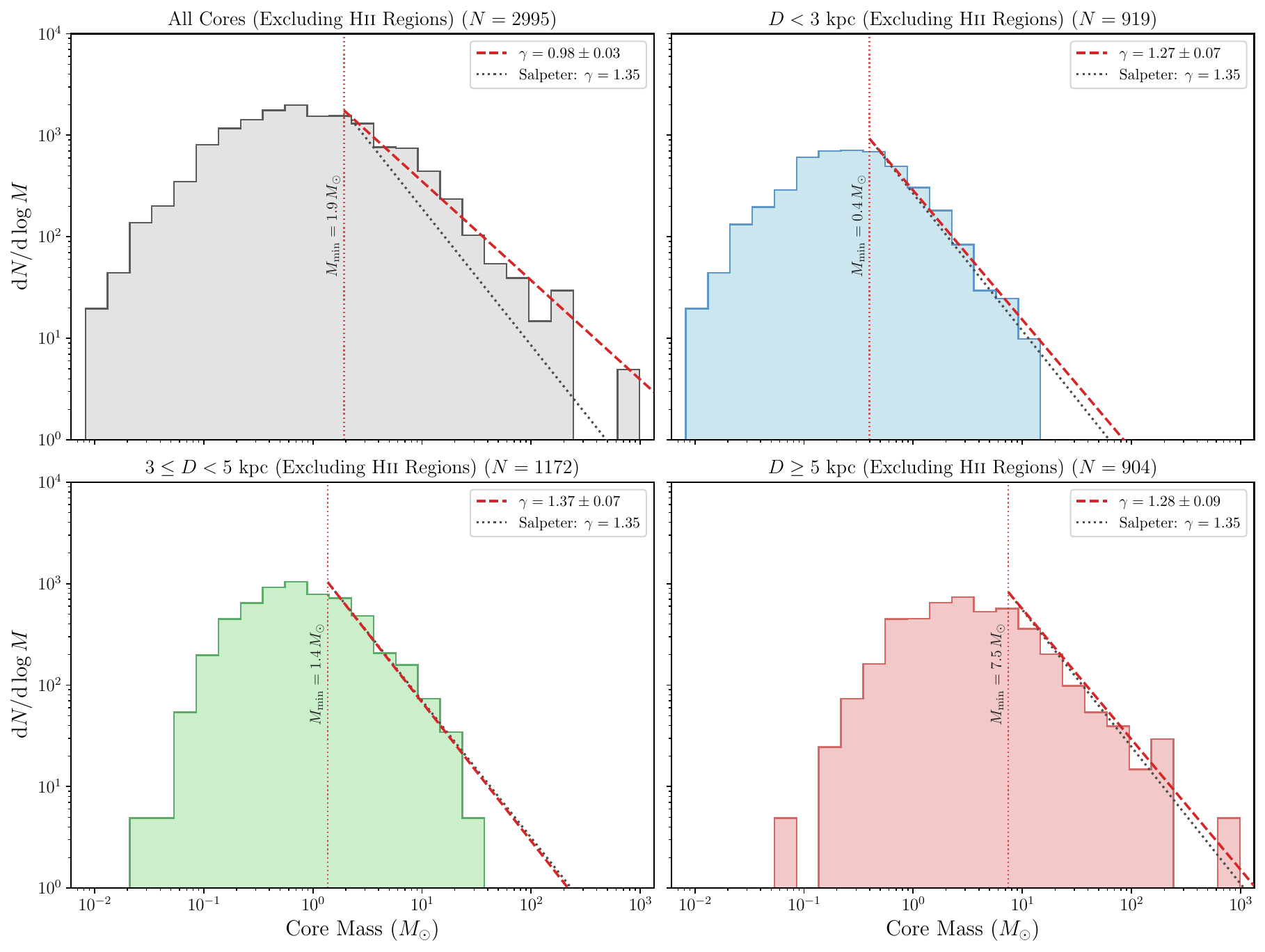}
    \caption{
\textbf{Top-left}: CMF of all dense cores excluding H{\sc ii} regions in the QUARKS sample. The vertical red dotted line marks the fitted lower mass limit, $M_{\min}$, whose value is labeled alongside the line. The red dashed line represents the best-fit power-law slope obtained using the unbinned MLE. The gray dotted line indicates the Salpeter slope, $\gamma=1.35$, for comparison. \textbf{Top-right}: CMF of all dense cores excluding H{\sc ii} regions in the near-distance subsample ($D<3$~kpc). \textbf{Bottom-left}: CMF of all dense cores excluding H{\sc ii} regions in the intermediate-distance subsample ($3 \leq D < 5$~kpc). \textbf{Bottom-right}: CMF of all dense cores excluding H{\sc ii} regions in the far-distance subsample ($D\geq5$~kpc).
}
\label{fig:cmf_all_cores_no_hii}
\end{figure*}

Dense cores represent the precursors of stars, and their mass distribution---the core mass function (CMF)---is considered to be linked to the IMF. Therefore, investigating the CMF in protoclusters provides important
insights into the origin of the IMF and the physics of star formation. In nearby low-mass star-forming regions, numerous studies have shown that the CMF often has a high-mass power-law slope similar to that of the IMF but is shifted toward higher masses (e.g., \citealt{Alves_2007,Andre_2010,Konyves_2015}). This similarity was interpreted as evidence that stellar masses are largely determined by the masses of their natal cores  and that fragmentation may play an important role in determining the IMF. However, observations of some massive star-forming regions have revealed CMFs that are more top-heavy than those measured in low-mass regions, with a relative excess of high-mass cores (e.g., \citealt{Li_2007,Motte_2018b,Lu_2020,Louvet_2024_ALMAIMF,Morii_2026}), suggesting that the measured CMF may vary among different star-forming environments.

The QUARKS sample consists of 139 active protoclusters, which allows us to systematically investigate the shape of the CMF in massive protocluster environments. As discussed in Sect.~\ref{sec:physical_property}, the masses of H{\sc ii} regions may be overestimated due to contamination from free--free emission \citep{Nony_2024}. We therefore fitted the global CMF of the QUARKS survey after excluding H{\sc ii} regions. The high-mass tail of the CMF was fitted following the methods described in \citet{Clauset_2009}. We assumed that, above a lower mass limit $M_{\min}$, the CMF can be described by a power-law distribution,
\begin{equation}
\frac{\mathrm{d}N}{\mathrm{d}M} \propto M^{-\alpha}, \quad M \ge M_{\min},
\end{equation}
which corresponds to $\mathrm{d}N/\mathrm{d}\log M \propto M^{-\gamma}$ with $\gamma = \alpha - 1$.

For a given trial value of $M_{\min}$, the power-law index $\alpha$ was estimated using the unbinned maximum-likelihood estimator (MLE) for a power-law distribution. The optimal lower mass limit $M_{\min}$ was determined by minimizing the Kolmogorov--Smirnov (KS) distance between the cumulative distribution of the observed core masses with $M \ge M_{\min}$ and that of the corresponding power-law model \citep{Clauset_2009}. Only values of $M_{\min}$ with a sufficient number of cores above the threshold were considered to ensure statistical robustness. The uncertainty on the fitted power-law index was calculated as
\begin{equation}
    \sigma(\alpha) \simeq \frac{\alpha - 1}{\sqrt{n_{\ge}}} + \mathcal{O}\!\left(\frac{1}{n_{\ge}}\right)
\end{equation}
where $n_{\ge}$ is the number of cores with $M \ge M_{\min}$.

Fig.~\ref{fig:cmf_all_cores_no_hii} shows the fitted global CMF of the QUARKS sample after excluding H{\sc ii} regions. For display purposes, the CMF is shown in binned form as $\mathrm{d}N/\mathrm{d}\log M$, while the power-law index is derived from the unbinned MLE. Under the adopted dust temperature assumptions, the observed global CMF has a high-mass slope of $\gamma=0.98\pm0.03$. This relatively flat slope describes the combined core population of the QUARKS survey and should not be interpreted directly as the intrinsic CMF of a typical individual protocluster.

The QUARKS sample spans a wide range of distances. For a fixed angular resolution, the physical resolution therefore varies substantially among the targets, leading to different levels of source blending and mass sensitivity. The relation between core mass and distance, together with the cumulative distance distributions of the four evolutionary stages, is shown in Fig.~\ref{fig:mass_distance_ecdf}. To examine how these effects may influence the measured CMF, we divided the QUARKS protoclusters into three distance bins: $D<3$ kpc, $3\leq D<5$ kpc, and $D\geq5$ kpc, containing 919, 1172, and 904 cores, respectively, after excluding H{\sc ii} region candidates. We then re-derived the CMF in each bin using the same fitting method as for the full sample. Under the adopted dust temperature assumptions, the measured high-mass slopes in the three distance bins are approximately Salpeter-like, with $\gamma=1.27\pm0.07$, $1.37\pm0.07$, and $1.28\pm0.09$ in order of increasing distance, as shown in Fig.~\ref{fig:cmf_all_cores_no_hii}. In contrast, the corresponding fitted lower mass limits increase systematically from $0.40\,M_\odot$ to $1.37\,M_\odot$ and then to $7.46\,M_\odot$. 
These indicate that the underlying CMF in massive protoclusters could be Salpeter-like at various physical scales (or fragmentation levels).

We further tested whether combining subsamples with Salpeter-like CMFs but different lower mass limits can lead to a flatter aggregated CMF. As an illustrative test, we constructed a simple toy model in which three Salpeter-like input CMFs are subject to distance-dependent selection functions. The details are described in Appendix~\ref{app:cmf_selection_toy}, and the results are shown in Fig.~\ref{fig:cmf_distance_selection_toy}. The aggregated mock CMF has $\gamma_{\rm toy}=1.04^{+0.06}_{-0.09}$, consistent with the observed global value. This simple test illustrates that the underlying CMF within massive protoclusters could be Salpeter-like, but the distance-dependent selection effects in observations with a uniform angular resolution can substantially change the slope of an aggregated CMF. 

Although dividing the sample into three distance bins reduces the variations in physical resolution, mass sensitivity, and source blending within each subsample, it does not fully remove distance bias or provide a completeness correction. 
The approximately Salpeter-like slopes measured in the three bins therefore do not fully demonstrate that the intrinsic CMF is necessarily Salpeter-like. 
Incompleteness at low masses, spatial filtering of extended emission, dust temperature uncertainties, and projection effects may also affect the measured CMF (e.g., \citealt{Louvet_2021,Padoan_2023,Zhang_2025}). These effects should be carefully assessed before an aggregated CMF is interpreted physically or linked to the stellar IMF. A more detailed CMF analysis will be carried out in future work.

\subsection{Core Formation Efficiency in Protoclusters}

Star formation activity in molecular clouds is inefficient, with only a small fraction of the available gas forming stars (e.g., \citealp{Krumholz_2007, Evans_2009}). Beyond the core-to-star conversion efficiency, another key step in this process is the conversion of clump-scale gas into dense cores, which represent the immediate sites of star formation. Therefore, the relationship between the total mass of dense cores and the mass of their parental clumps provides a fundamental observational constraint on how effectively gas is assembled into star-forming substructures \citep{Bontemps_2010}, and previous studies have shown that the mass contained in dense cores is closely linked to the global properties of their parent clumps, suggesting that the efficiency of mass concentration is regulated by large-scale physical conditions (e.g., \citealt{Louvet_2014,Csengeri_2017,Traficante_2023,Xu_2024_ASSEMBLE}).

\begin{figure*}
  \centering
  \includegraphics[width=0.98\textwidth]{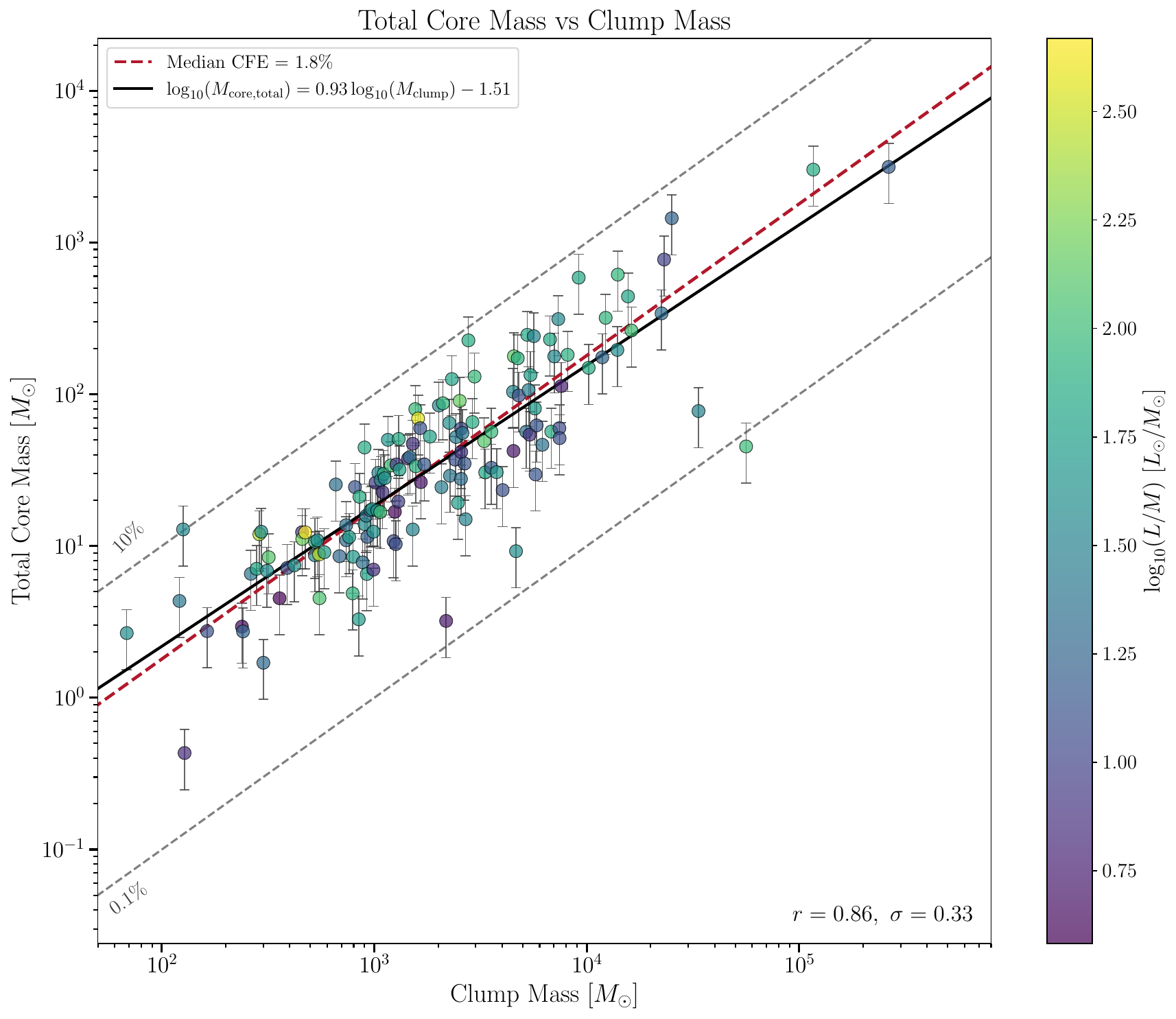}
  \caption{Relation between the total core mass and the parental clump mass in the QUARKS sample. Each colored data point represents one clump, with the color representing its luminosity-to-mass ratio. The parental clump masses and luminosity-to-mass ratios are adopted from \citet{Liu_2024_QUARKSI}. The gray dashed lines indicate constant core-to-clump mass fractions of 10\% and 0.1\%. The red dashed line marks the median core formation efficiency of the sample, 1.8\%. The black solid line shows a linear regression in logarithmic space, yielding the scaling relation $\log_{10} (M_{\rm core,total}) = 0.93 \log_{10} (M_{\rm clump}) - 1.51$, with a Pearson correlation coefficient of $r = 0.86$ and a $1\sigma$ scatter of 0.33 dex. The error bars show the propagated uncertainties in the total core masses, based on the uncertainty estimates described in Sect.~\ref{sec:physical_property}.
  }
  \label{fig:core_clump_mass}
\end{figure*}

Using the QUARKS survey, we investigate this mass redistribution by examining the relation between the total core mass and the clump mass. As shown in Fig.~\ref{fig:core_clump_mass}, these two parameters exhibit a tight linear correlation with a best-fit slope close to unity, indicating that the core formation efficiency (CFE, $M_{\rm{core, total}}/M_{\rm{clump}}$) is approximately uniform across different protoclusters within the QUARKS sample, despite moderate scatter. Only a small fraction of the clump mass--a median of approximately 2\%--is concentrated into compact dense cores.

While the CFE remains internally consistent within the QUARKS sample, its absolute value is highly sensitive to the observational setup, including spatial resolution, maximum recoverable scale, and sensitivity. For instance, compared with the results reported in \citet{Xu_2024_Quarks}, adopting a different array configuration causes the median CFE to shift from $3.9\%$ to $1.8\%$. Consequently, absolute CFE values should not be directly compared between different projects unless differences in spatial resolution, spatial filtering of extended emission, sensitivity, and analysis methods are carefully accounted for.

\begin{figure}
  \centering
  \includegraphics[width=\columnwidth]{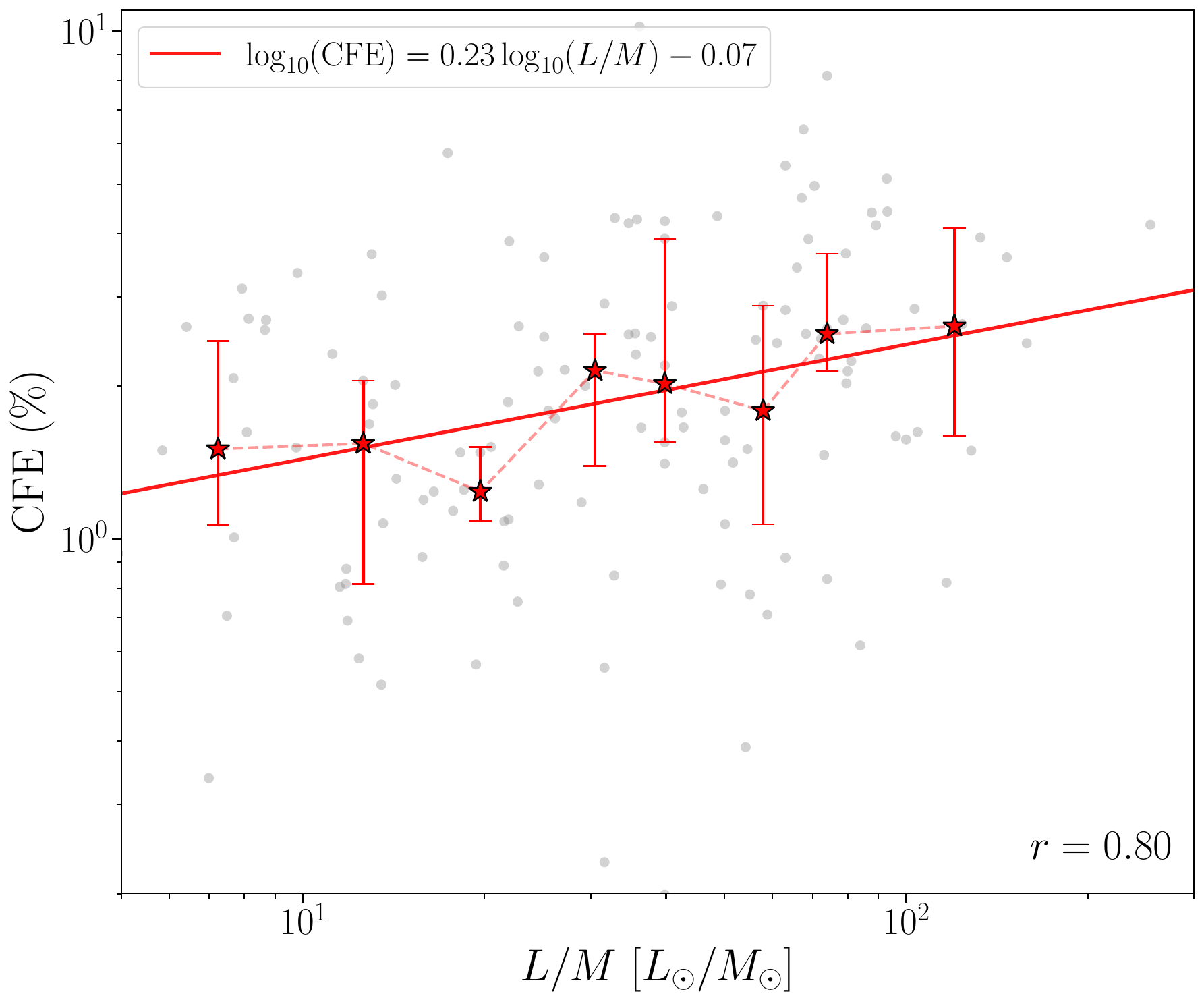}
  \caption{The relation between the CFE and the clump luminosity-to-mass ratio. The clump luminosity-to-mass ratios are adopted from \citet{Liu_2024_QUARKSI}. Grey points show individual clumps, while the red stars indicate the median CFE in equal-number bins of $\log(L/M)$; the error bars represent the interquartile range (25th--75th percentiles). The solid red line represents a linear fit to the binned medians in logarithmic space, and the Pearson correlation coefficient calculated from the binned medians is labeled in the bottom-right corner.
  }
  \label{fig:cfe_lm}
\end{figure}

We further investigated whether parental clump properties influence the CFE. While no significant correlation was found with clump-averaged density or source distance (Fig.~\ref{fig:CFE_VS_parameter}), the CFE exhibits a positive correlation with the luminosity-to-mass ratio ($L/M$; Fig.~\ref{fig:cfe_lm}). Clumps with higher $L/M$ tend to have systematically higher median CFEs, suggesting that clump evolution is accompanied by an increasing fraction of mass being concentrated into dense cores. This trend, despite its substantial scatter, is consistent with the Rosetta Stone simulations, in which $L/M$ increases with star formation efficiency as embedded protostars form, accrete, and progressively dominate the luminosity of the clump \citep{Tung2025, Nucara_2025}. A similar increase in CFE with $L/M$ has also been reported observationally by the ALMAGAL survey \citep{Elia_2026} and across the evolutionary sequence examined by \citet{Xu_2026}. In this picture, the observed CFE--$L/M$ correlation reflects the coupled growth of the dense-core population and the embedded protostellar population during clump evolution, further supporting the ``clump-fed'' scenario of star formation from an observational perspective. Feedback-driven dispersal of lower-density clump gas may also affect the clump mass and the derived CFE.

\begin{figure*}[t]
    \centering
    \includegraphics[width=\textwidth]{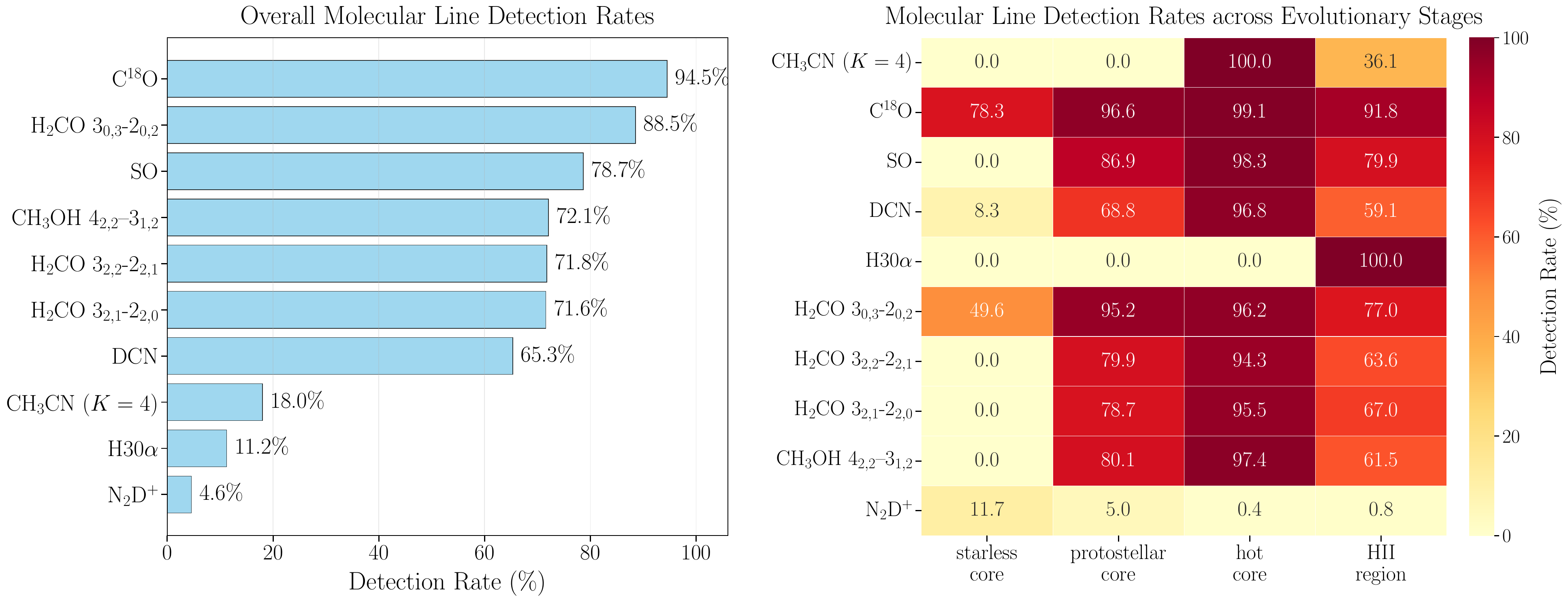}
    \caption{\textbf{Left:} Overall detection rates of the observed molecular transitions toward dense cores in the QUARKS survey. \textbf{Right:} Detection-rate heatmap of molecular lines across evolutionary stages, highlighting systematic chemical variations during core evolution.}
\label{fig:detection_rates}
\end{figure*}

\subsection{Core Chemical Evolution}

Molecular lines are powerful tracers of the chemical and physical processes associated with star formation and core evolution, as different species are preferentially excited or enhanced under specific environmental conditions (e.g., \citealt{Bergin_2007, Caselli_2012}). In particular, the deuterium fraction and the detection rates of commonly observed molecular transitions provide valuable diagnostics for characterizing the chemical state of dense cores and their evolution over time (e.g., \citealt{Fontani_2011, Sanhueza_2012, Kim_2020, Li_2022}). As dense cores evolve from cold prestellar objects to actively accreting protostellar systems, their increasing temperatures and the release of molecules from dust grains significantly reshape their chemical composition. Molecules that are efficiently formed under cold and CO-depleted conditions, such as deuterated species, tend to decline as protostellar heating becomes significant, while warm gas tracers and complex organic molecules become more easily detectable \citep{Aikawa_2008, Jorgensen_2020}.

In this work, we investigate the detection rates of common molecular tracers within dense cores at different evolutionary stages using the QUARKS survey data. Fig.~\ref{fig:detection_rates} summarizes the detection statistics of the observed transitions. The left panel presents the overall detection rates toward the full core sample. Several diffuse-gas tracers are widely detected, while warm-gas tracers (e.g., $\rm{CH_3OH}$ and $\rm{H_2CO}$ transitions) show detection rates of nearly 70\%. In contrast, the high-excitation $\rm{CH_3CN}$ line, the H30$\alpha$ radio recombination line, and the cold-gas tracer $\rm{N_2D^+}$ are detected far less frequently. The right panel further illustrates the detection-rate heatmap across evolutionary stages and reveals clear chemical differentiation among molecular tracers. Some detection-rate differences are expected from the classification criteria, since several of these transitions are used to identify the core evolutionary stages. The overall detection patterns suggest increasing chemical complexity as dense cores evolve from prestellar to hot-core phases, followed by a potential decline after the formation of H{\sc ii} regions due to strong stellar feedback. Notably, $\rm{N_2D^+}$ exhibits the opposite trend, with higher detection rates in earlier stages, suggesting that it serves as an effective tracer of cold and young cores.

Previous chemical models have shown that deuterated molecules are highly sensitive to environmental conditions such as temperature, density, and CO depletion \citep{Caselli_2002, Ceccarelli_2014}. In cold and well-shielded regions, enhanced deuteration is expected, whereas increasing temperatures and protostellar activity reshape the deuterium chemistry. Observational studies further demonstrate that the evolutionary behavior of deuteration is molecule-dependent: the deuterium fraction of $\rm{N_2H^+}$ typically declines as cores evolve \citep[e.g.,][]{Fontani_2011, Kong_2016}, while species such as HCN and HNC can exhibit higher deuteration during the protostellar phase \citep[e.g.,][]{Gerner_2015, Taniguchi_2024}. Consequently, deuterated species are widely regarded as effective chemical clocks, and a multi-species view of deuteration offers a more comprehensive probe of core evolution than a single molecular tracer \citep[e.g.,][]{Sakai_2022, Jiao_2023}.

\begin{figure*}[t]
\centering
\includegraphics[width=\textwidth]{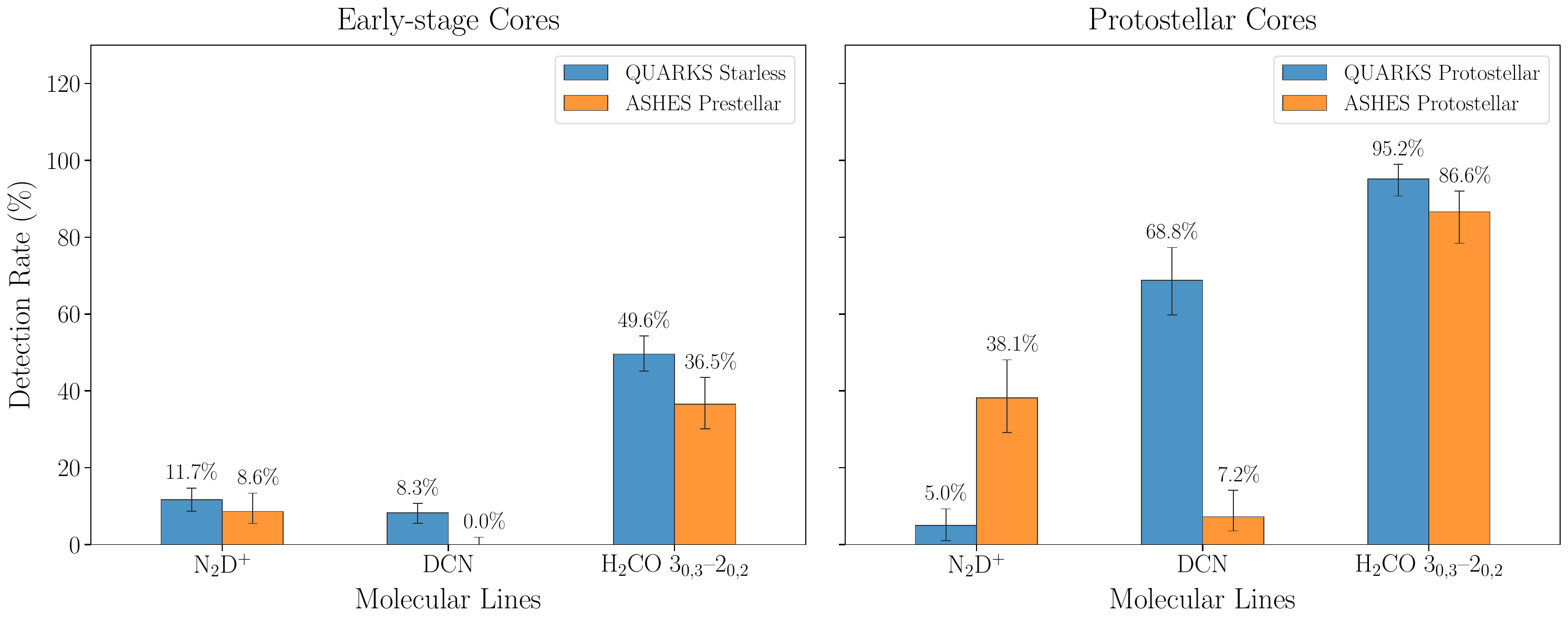}
\caption{Comparison of deuterated molecular-line detection rates between the QUARKS and ASHES surveys \citep{Li_2022} for early-stage and protostellar cores. \textbf{Left}: Detection rates toward early-stage cores. The H$_2$CO line is included as a reference tracer to characterize the overall chemical properties of the parental clumps. The starless core classification adopted in QUARKS is consistent with the prestellar core definition used in ASHES. \textbf{Right}: Detection rates toward protostellar cores in both surveys. The QUARKS error bars show the central 95\% ranges obtained from $10^4$ random subsamples containing the same numbers of cores as the corresponding ASHES samples. The ASHES error bars represent the 95\% Wilson confidence intervals for the observed detection fractions.}

\label{fig:quarks_ashes_comparison}
\end{figure*}

By comparing the detection rates of deuterated species in dense cores at matched evolutionary stages in the QUARKS and ASHES (The ALMA Survey of 70~$\mu$m dark High-mass clumps in Early Stages) surveys, we assess how clump-scale environments affect core chemistry. Fig.~\ref{fig:quarks_ashes_comparison} shows that starless cores in QUARKS have slightly higher detection rates than those in ASHES, although the differences are small. In contrast, a clear divergence emerges in the protostellar phase: $\mathrm{N_2D^+}$ is detected about eight times more frequently in ASHES than in QUARKS, whereas DCN is detected about ten times more frequently in QUARKS than in ASHES. This difference likely reflects both environmental effects and sample selection. ASHES targets 70~$\mu$m-dark high-mass clumps, which are generally colder and chemically younger, providing favorable conditions for CO depletion and enhanced $\mathrm{N_2D^+}$ emission. In contrast, QUARKS targets active protoclusters selected from the ATOMS sample, where stronger internal heating, feedback, and more complex clustered environments can reduce the abundance of $\mathrm{N_2D^+}$ while enhancing warmer deuterated tracers such as DCN. These contrasting trends suggest that the differences in environmental conditions, stellar feedback and sample selection can strongly influence the chemical composition of dense cores, so that even cores classified at similar evolutionary stages may show different deuteration patterns.

\section{Summary and  Conclusions} \label{sec:conclusions}

We present a systematic analysis of dense cores identified in the QUARKS survey, which targets 139 high-mass star-forming clumps observed with ALMA at 1.3 mm with an angular resolution of approximately $0.3^{\prime\prime}$, corresponding to a physical scale of $\sim1160$ au at a median distance of 3.8 kpc. Using the \texttt{getsf} algorithm for core extraction, we constructed a homogeneous catalog of dense cores in massive protocluster environments, enabling statistical studies of the core mass function (CMF), core formation efficiency (CFE), and chemical evolution. Our main conclusions are as follows:

\begin{enumerate}

\item We identified 3374 dense cores from the continuum images and classified them using molecular-line diagnostics and outflow associations, resulting in 351 starless core candidates ($10.4\%$), 2174 protostellar core candidates ($64.4\%$), 470 hot core candidates ($13.9\%$), and 379 H{\sc ii} region candidates ($11.2\%$). Under the adopted dust temperature assumptions, the core mass shows a statistically significant increasing trend along the evolutionary stages, whereas the molecular hydrogen number density does not show a clear trend.

\item  Under the adopted dust temperature assumptions, the observed global CMF after excluding H{\sc ii} region candidates has a relatively flat high-mass slope of $\gamma=0.98\pm0.03$. In contrast, the high-mass slopes determined in the three distance bins are approximately Salpeter-like, while their fitted lower mass limits increase systematically with distance. The difference between the distance-binned and aggregated results, together with the illustrative toy model, shows that distance-dependent observational selection can substantially affect the slope of an aggregated CMF. Therefore, future studies of the CMF should carefully assess and account for observational effects before interpreting its shape or linking it to the stellar IMF.

\item  We measured the CFE in active massive protoclusters and found that a median of approximately $2\%$ of the clump mass is concentrated in dense cores, with relatively little variation across diverse physical environments. We further examined its dependence on the luminosity-to-mass ratio and identified a positive correlation between CFE and $L/M$. Taken together, these results suggest that dense core formation maintains a relatively steady efficiency that gradually increases as clumps evolve. This trend is consistent with ongoing gas accretion in dynamically evolving protoclusters. 

\item We analyzed the molecular-line detection rates of dense cores across evolutionary stages and found that line richness increases from the earliest stages to the hot core phase, where it reaches a maximum. We further compared the detection rates of deuterated species with those reported in the ASHES survey and found that protostellar cores in QUARKS exhibit lower N$_2$D$^+$ and higher DCN detection rates. These systematic differences suggest that parental-clump environments can influence the chemical evolution of dense cores.

\end{enumerate}

Overall, the QUARKS core catalog provides one of the largest samples of dense cores in massive star-forming clumps at 1.3 mm, offering a robust statistical framework for probing the early stages of protocluster formation. The observed relations between core and clump properties are consistent with a clump-fed scenario in active massive protoclusters. Future joint ATOMS--QUARKS studies, combined with multiwavelength observations, higher-resolution data, and numerical simulations, will further constrain the physical properties of dense cores and their interactions with the surrounding environment.

\begin{acknowledgments}
This work has been supported by the National Science and Technology Major Project of China (No. 2024ZD1100601) and National Key R\&D Program of China (No. 2022YFA1603100).  W.J. acknowledges support from the Shanghai Post-doctoral Excellence Program. T.L. acknowledges support from the National Natural Science Foundation of China (NSFC), through grants No. 12073061 and No. 12122307, the Tianchi Talent Program of Xin-jiang Uygur Autonomous Region, and the international partnership program of the Chinese Academy of Sciences, through grant No. 114231KYSB20200009. 
PS was partially supported by a Grant-in-Aid for Scientific Research (KAKENHI Number JP26H02066 and JP23H01221) of JSPS. G.G. acknowledges support by the ANID BASAL project FB210003. MJ acknowledges the support of the Research Council of Finland Grant No.
348342. S.L. acknowledges support from the National SKA Program of China with No. 2025SKA0140100, “Double First-Class” Funding with No. 14912217, and National Natural Science Foundation of China (NSFC) grant with No. 13004007. This research was carried out in part at the Jet Propulsion Laboratory, which is operated by the California Institute of Technology under a contract with the National Aeronautics and Space Administration (80NM0018D0004). SRD acknowledges support from the Fondecyt Postdoctoral fellowship (project code 3220162) and ANID BASAL project FB210003. A.S. is supported by the China Postdoctoral Science Foundation through the grant No. 2026M793499. This paper makes use of the following ALMA data: ADS/JAO.ALMA\#2021.1.00095.S. ALMA is a partnership of ESO (representing its member states), NSF (USA) and NINS (Japan), together with NRC (Canada), NSTC and ASIAA (Taiwan), and KASI (Republic of Korea), in cooperation with the Republic of Chile. The Joint ALMA Observatory is operated by ESO, AUI/NRAO and NAOJ. The National Radio Astronomy Observatory is a facility of the National Science Foundation operated under cooperative agreement by Associated Universities, Inc.
\end{acknowledgments}





%
\facilities{Spitzer, APEX, MeerKAT, ALMA}

\software{astropy \citep{2013A&A...558A..33A,2018AJ....156..123A,2022ApJ...935..167A}, \texttt{getsf} \citep{Menshchikov_2021_getsf}
          }


\appendix

\section{TM1 Observational Information}
\label{app:TM1_obs}
\setcounter{table}{0}
\renewcommand{\thetable}{A\arabic{table}}
This appendix summarizes the observational setup of the QUARKS TM1 data. Table~\ref{tab:TM1_obs} lists the scheduling block information, including the Group ID, number of fields, source name, observing date, number of antennas,  average integration time per pointing, baseline range, angular resolution, maximum recoverable scale, and the adopted calibrators.

\begin{table*}[!thb]
\centering
\caption{QUARKS TM1 Observation Information. \label{tab:TM1_obs}}
\renewcommand{\arraystretch}{1.2} 
\begin{tabular}{ccccccccccc}
\hline
\hline
$\rm Group \ ID$ & $\rm n_{field}$ & Source & Obs. Date &Antenna &  Int. Time& Min./Max. BL & AR & MRS & \multicolumn{2}{c}{Calibrators}  \\
\cline{10-11}
& & & & & (s) & (m/m) & (\arcsec) & (\arcsec) & Phase & Bandpass/Flux  \\
(1) & (2) & (3) & (4) & (5) & (6) & (7) & (8)&(9) & (10) & (11) \\
\hline
1 & 7& 08303-09094 & 2023-04-08 & 44 & 242 &15.3/1397.4 & 0.2 & 3.4 & J0922-3959 & J1107-4449  \\
\hline
2 & 4 &10365-11332& 2024-05-31 & 42& 272 &15.1/1397.8 & 0.2 & 3.5 & J1047-6217 & J1107-4449  \\
\hline
\multirow{2}{*}{3} & \multirow{2}{*}{15} &\multirow{2}{*}{12320-13295}& 2023-05-02 & 46 &  152 &15.3/2516.8 & 0.1 & 2.2 & J1337-6509 & J1617-5848  \\ 
 &  && 2024-06-02 & 40 & 151 &15.1/1397.8 & 0.2 & 3.5 & J1337-6509 & J1617-5848  \\
\hline
4 & 8 &13471-14212& 2024-06-04 & 42 & 272 &15.1/1397.8 & 0.2 & 3.8 & J1408-5712 & J1617-5848  \\
\hline
5 & 5 &14382-15290& 2024-06-02 & 41 & 267 &15.1/1397.8 & 0.2 & 3.5 & J1524-5903 & J1617-5848  \\
\hline

 &  &  & 2024-06-02 & 41 & 91 &15.1/1397.8 & 0.2 & 3.5 & J1603-4904 & J1617-5848  \\    
6 &  28 & 15384-16177 & 2024-06-03 & 45 & 91 &15.1/1397.8 & 0.3 & 3.8 & J1603-4904 & J1617-5848  \\
  &  && 2024-06-03 & 45 & 16 &15.1/1397.8 & 0.3 & 3.8 & J1603-4904 & J1617-5848  \\
\hline

 &  &  & 2024-06-02 & 41 & 113 &15.1/1397.8 & 0.2 & 3.5 & J1650-5044 & J1617-5848  \\
 7&  19 & 16272-16489 & 2024-06-03 & 45 & 97 &15.1/1397.8 & 0.3 & 3.8 & J1650-5044 & J1617-5848  \\
  &  && 2024-06-03 & 44 & 121 &15.1/1397.8 & 0.2 & 4.0 & J1650-5044 & J1617-5848  \\
\hline
8 & 7 &16524-17016& 2022-08-12 & 46 & 242 &15.1/1301.6 & 0.3 & 3.5 & J1636-4102 & J1617-5848  \\
\hline
\multirow{2}{*}{9} & \multirow{2}{*}{15} &\multirow{2}{*}{17136-17278}& 2022-08-16 & 42& 121 &15.1/1301.6 & 0.2 & 3.7 & J1733-3722 & J1617-5848  \\
&  && 2022-08-18 & 44 & 121 &15.1/1301.6 & 0.3 & 3.8 & J1733-3722 & J1617-5848  \\
\hline
10 & 4 &17439-17455& 2022-08-13 & 47 & 227 &15.1/1301.6 & 0.3 & 3.5 & J1744-3116 & J1924-2914  \\   \hline
11 & 6 &17545-18075& 2022-08-13 & 47 & 212 &15.1/1301.6 & 0.3 & 3.5 & J1751-1950 & J1924-2914  \\   \hline
12 & 13 &18079-18264& 2022-08-14 & 46&  216 & 15.1/1301.6 & 0.3 & 3.5 & J1832-2039 & J1924-2914  \\    \hline
13 & 11 &18290-18341& 2022-08-16 & 41 & 242 & 15.1/1301.6 & 0.2 & 3.7 & J1832-1035 & J1924-2914  \\     \hline
14 & 11 &18411-18530& 2022-08-15 & 46 & 225 &15.1/1301.6 & 0.3 & 3.5 & J1851+0035 & J1924-2914  \\  
\hline
15 & 3 &19078-19097& 2022-08-15 & 45 & 242 &15.1/1301.6 & 0.3 & 3.5 & J1905+0952 & J1924-2914  \\
\hline
\end{tabular}{
\begin{flushleft}  
The QUARKS observations were arranged into 15 SBs (scheduling blocks), which are referred to as “Groups” here. The Group ID is listed in Column 1. The number of single-pointing fields in each group is listed in Column 2. The sources contained in each group are listed in Column 3. The observation date, number of antennas used, and average integration time per pointing for each observation are listed in Columns 4–6, respectively. The minimum and maximum baselines (BL) are listed in Column 7. The angular resolution (AR) and maximum recoverable scale (MRS) are listed in Columns 8–9. The phase calibrator and the bandpass/flux calibrators are listed in Columns 10–11.

\end{flushleft}
}
\end{table*}

\section{Overview of Compact-Source Extraction Algorithms}
\label{app:source_extraction}

Several algorithms have been developed to identify compact sources in continuum images. Their different treatments of source boundaries, hierarchical structures, and diffuse backgrounds can lead to differences in the number, morphology, and measured fluxes of extracted sources. Here, we briefly summarize four algorithms commonly used in dense-core studies.

\texttt{CLUMPFIND} \citep{Williams_1994_clumpfind} identifies structures by dividing an emission map into a series of intensity contours and assigning connected pixels to local maxima. It has been widely applied to continuum observations of dense cores \citep[e.g.,][]{Enoch_2006,Di_francesco_2008}. The method is simple and intuitive, but its results can be sensitive to the adopted contour levels and may be affected by over- or under-segmentation.

\texttt{astrodendro} \citep{Rosolowsky_2008_astrodendro} describes emission using a hierarchical tree of trunks, branches, and leaves, allowing nested structures to be identified over multiple intensity levels. It has been extensively used to identify dense cores in high-mass star-forming regions \citep[e.g.,][]{Sanhueza_2019_ASHES,Coletta_2025_ALMAGAL}. However, because source boundaries are primarily defined by emission contours, the extracted structures can be irregular in regions containing extended emission. Flux measurements may also include contributions from the surrounding background unless an additional background-subtraction procedure is applied.

\texttt{CuTEx} \citep{Molinari_2011_cutex} uses second-derivative filtering to enhance compact sources against spatially varying backgrounds and has been applied to large continuum surveys \citep[e.g.,][]{Molinari_2016,Elia_2017_Higal}. It performs well for approximately Gaussian compact sources in crowded fields, but may be less sensitive to extended or irregular structures.

\texttt{getsf} \citep{Menshchikov_2021_getsf} uses a multiscale decomposition to separate compact sources, filaments, and diffuse backgrounds. It has been adopted in recent high-angular-resolution studies of dense cores \citep[e.g.,][]{Xu_2024_ASSEMBLE,Louvet_2024_ALMAIMF}. This approach allows sources over a range of spatial scales to be identified while accounting for variations in the surrounding emission. However, its background and filament decomposition may become uncertain in particularly complex regions, such as interferometric images affected by strong sidelobes, and the extraction procedure can be computationally intensive.

Several studies have compared the performance of these algorithms. \citet{Massi_2019} applied \texttt{CuTEx} and \texttt{CLUMPFIND} to dense cores in Vela~C and found that \texttt{CLUMPFIND} generally recovered a larger total core mass, whereas \texttt{CuTEx} was more sensitive to compact structures. In the ALMA-IMF survey, \citet{Louvet_2024_ALMAIMF} compared \texttt{getsf} with \texttt{GExt2D}, which uses a similar method to that of \textit{CuTEx}, and found that \texttt{getsf} produced more conservative extractions and more consistent source catalogs across images with different resolutions. Using ALMAGAL observations, \citet{Coletta_2025_ALMAGAL} compared \texttt{CuTEx} and \texttt{astrodendro} and found that \texttt{CuTEx} was more effective at identifying compact sources and measuring their fluxes.

\section{Distance-dependent Effects on the CMF Shape}
\label{app:cmf_selection_toy}

\begin{figure*}[t]
    \centering
    \includegraphics[width=\textwidth]{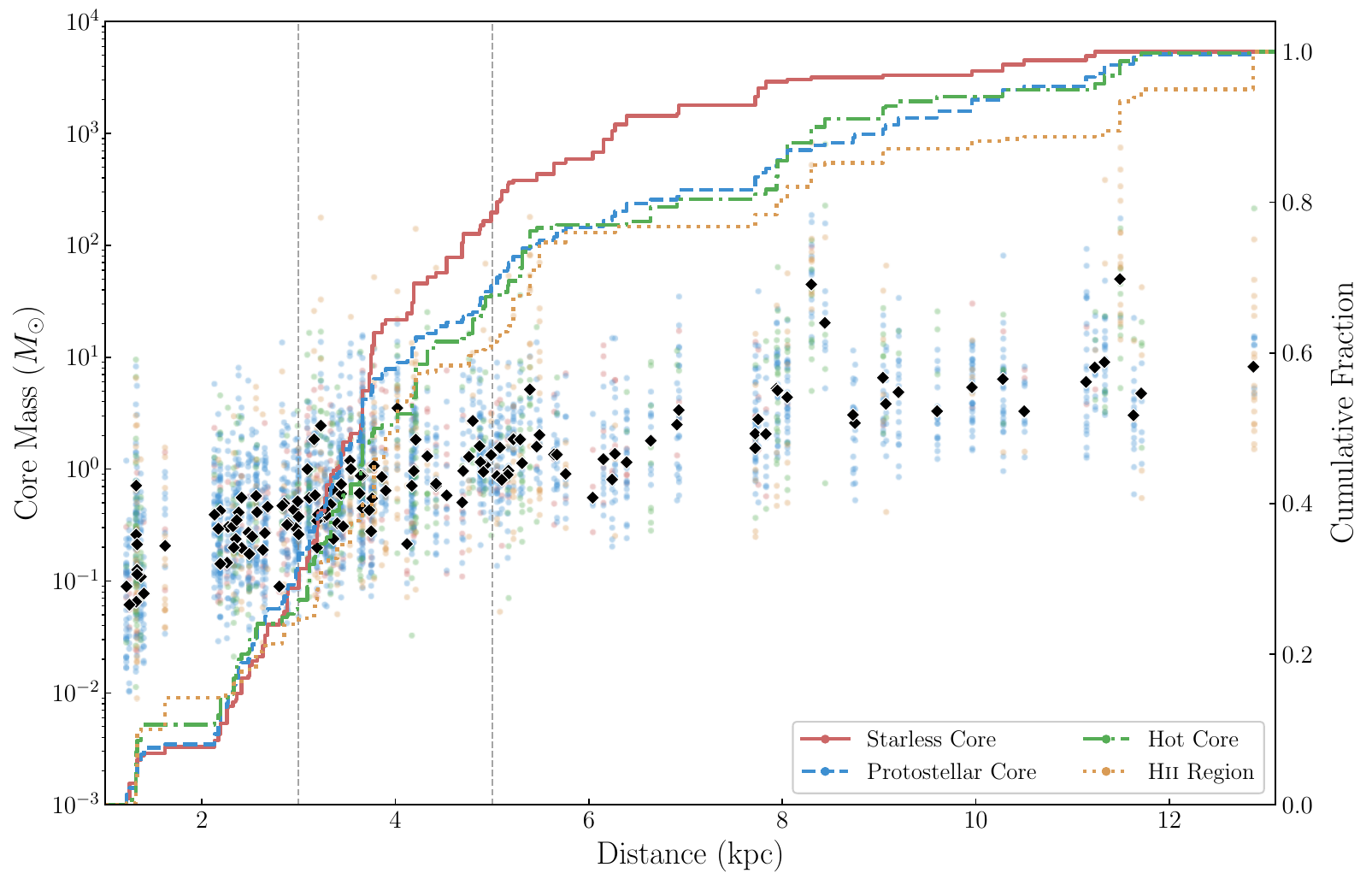}
    \caption{
Core masses as a function of distance and cumulative distance distributions of cores in the four evolutionary stages. Colored points represent individual core masses (left axis), with colors indicating evolutionary stages. Black diamonds mark the median mass of all cores within each parental clump. Colored step curves show the cumulative distribution function of distance for each evolutionary stage (right axis). Vertical dashed lines mark the boundaries of the adopted distance bins at 3 and 5 kpc.}
    \label{fig:mass_distance_ecdf}
\end{figure*}

Fig.~\ref{fig:mass_distance_ecdf} displays individual core masses as a function of distance, together with normalized cumulative distance distributions for the four evolutionary stages. While core masses exhibit a clear increasing trend with distance, this trend is considerably weaker within each individual distance bin.

To illustrate how distance-dependent observational effects can affect the aggregated CMF, we constructed a simple forward toy model in three distance bins: $D<3$ kpc, $3\leq D<5$ kpc, and $D\geq5$ kpc. The number of mock cores in each bin was fixed to that of the corresponding QUARKS subsample. The intrinsic CMF was assumed to follow
\begin{equation}
\frac{\mathrm{d}N}{\mathrm{d}\log M}\propto M^{-\gamma_{\rm input}},
\end{equation}
where a common slope was drawn from a Gaussian distribution, $\gamma_{\rm{input}}=1.35\pm0.05$, for the three distance bins. Thus, the input CMFs are Salpeter-like, with a small dispersion around the canonical slope.

We applied a gradual mass-dependent detection function to each distance bin,
\begin{equation}
C_i(M)=\frac{1}{2}\left[1+\mathrm{erf}
\left(\frac{\log M-\log M_{50,i}}{\sqrt{2}w_i}\right)\right],
\end{equation}
such that the observed mock CMF is proportional to
$M^{-\gamma_{\rm input}}C_i(M)$. Here, $M_{50,i}$ is the mass at which the relative detection probability reaches 50\%, and $w_i$ is the width of the transition in logarithmic mass. The fitted $w_i$ values are very similar and the main difference between the three mock CMFs is the value of $M_{50,i}$. We adopted $M_{50}=0.39$, 1.08, and
$5.02\,M_\odot$ for the near, intermediate, and far distance bins, respectively. These increasing turnover masses provide an illustrative representation of the distance-dependent observational effects expected from changing physical resolution and sensitivity.

\begin{figure*}[t]
    \centering
    \includegraphics[width=\textwidth]{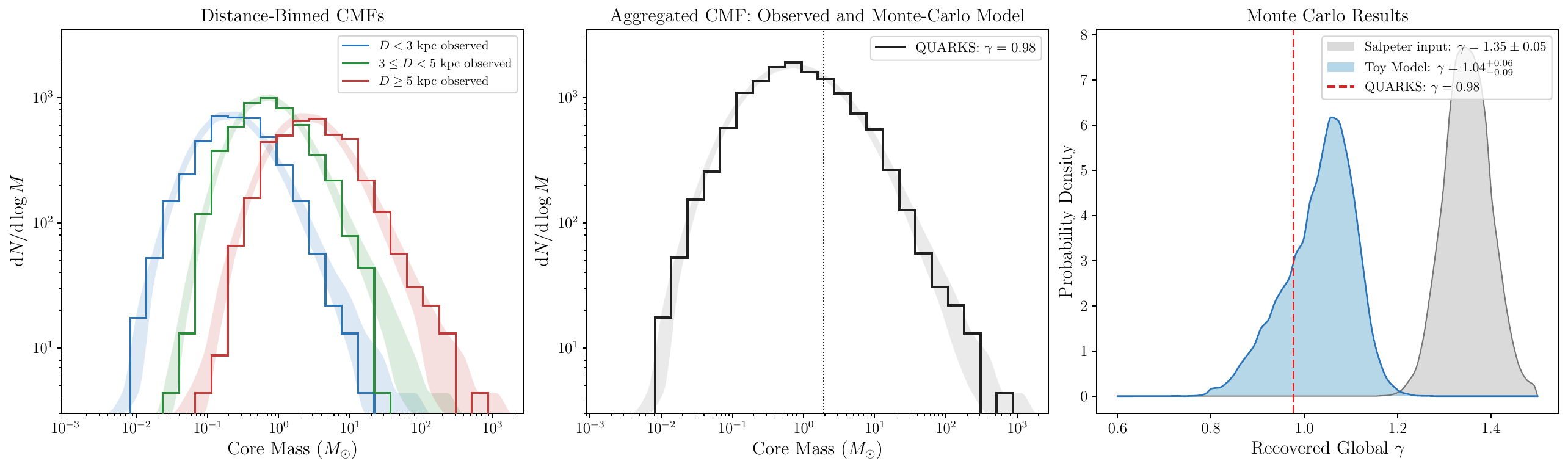}
    \caption{Illustrative toy model for the effect of distance-dependent
    selection on the aggregated CMF. \textbf{Left:} Observed distance-binned
    CMFs (colored histograms) and the 16--84\% ranges of the corresponding
    mock CMFs (shaded regions) from $10^4$ Monte Carlo simulations.
    \textbf{Middle:} Observed aggregated CMF (black histogram) and the
    16--84\% range of the aggregated mock CMFs (grey region).
    \textbf{Right:} Probability distributions of the Salpeter-like input slopes (grey) and the recovered slopes of the aggregated mock CMFs (blue). The red dashed line marks the observed global QUARKS slope.}
    \label{fig:cmf_distance_selection_toy}
\end{figure*}

We carried out $10^4$ Monte Carlo simulations. In each simulation, the three mock CMFs were combined and the aggregated CMF was fitted with the same unbinned MLE and KS procedure used for the QUARKS data. Although every input distance-bin CMF is Salpeter-like, their combination produces a systematically flatter aggregated CMF, as shown in Fig.~\ref{fig:cmf_distance_selection_toy}.
The recovered slope distribution has
$\gamma_{\rm toy}=1.04^{+0.06}_{-0.09}$, which can be compared directly with the observed global value of $\gamma=0.98\pm0.03$. The observed value lies within the 16--84\% range of the toy-model results.

This experiment is illustrative rather than a completeness correction: the
selection functions are calibrated from the observed CMF turnovers and are not
derived from injection--recovery tests. It does not demonstrate that the
observed global CMF slope is caused uniquely by distance-dependent resolution.
Instead, it shows that plausible distance-dependent turnover masses are
sufficient to transform the combination of three Salpeter-like CMFs into an
apparently top-heavy aggregated CMF.

\section{Additional Figures} \label{sec:appendix_a}



\begin{figure*}[t]
\centering

\begin{minipage}[t]{0.48\textwidth}
    \centering
    \includegraphics[width=\linewidth]{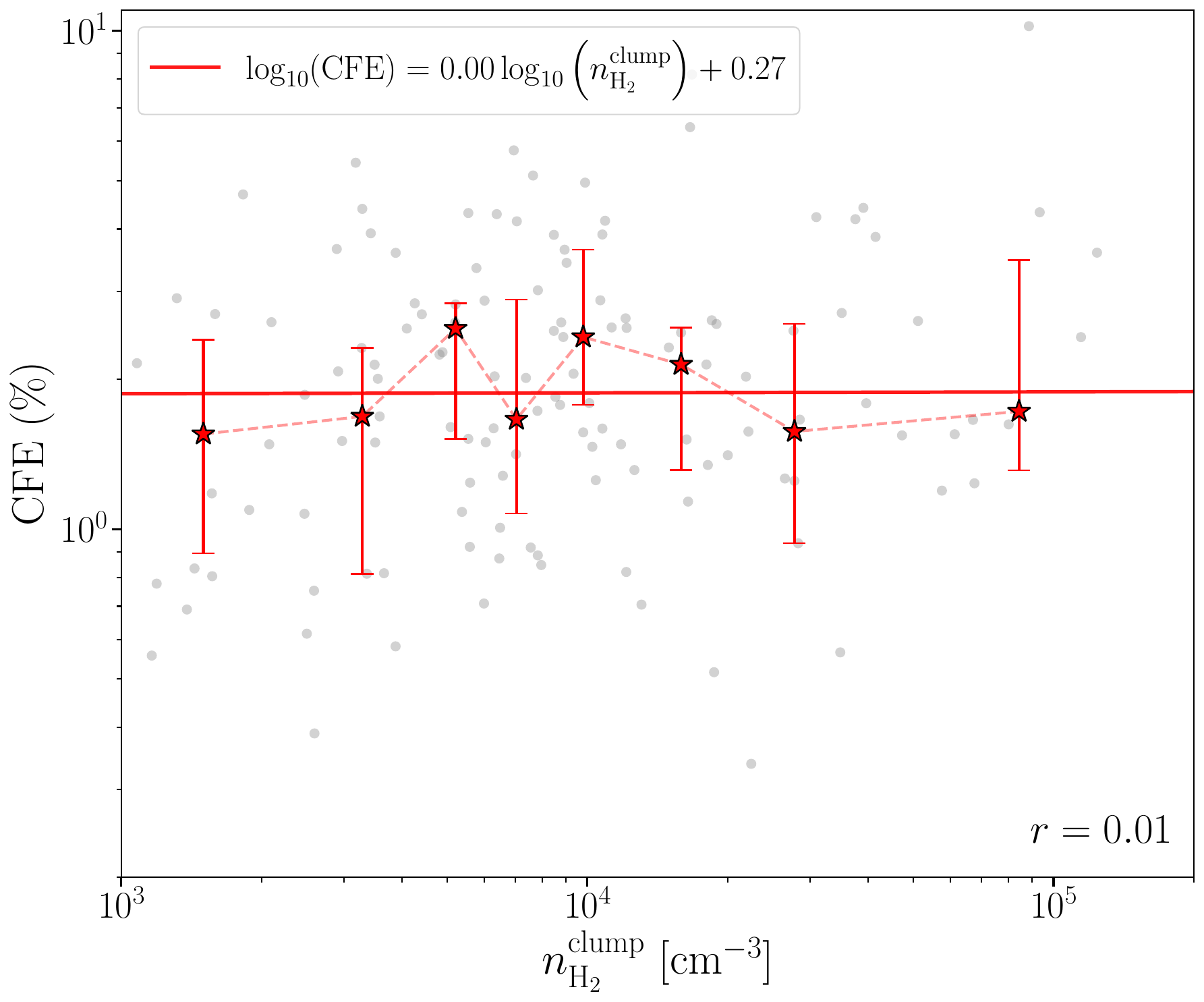}
    \vspace{-2mm}
\end{minipage}
\hfill
\begin{minipage}[t]{0.48\textwidth}
    \centering
    \includegraphics[width=\linewidth]{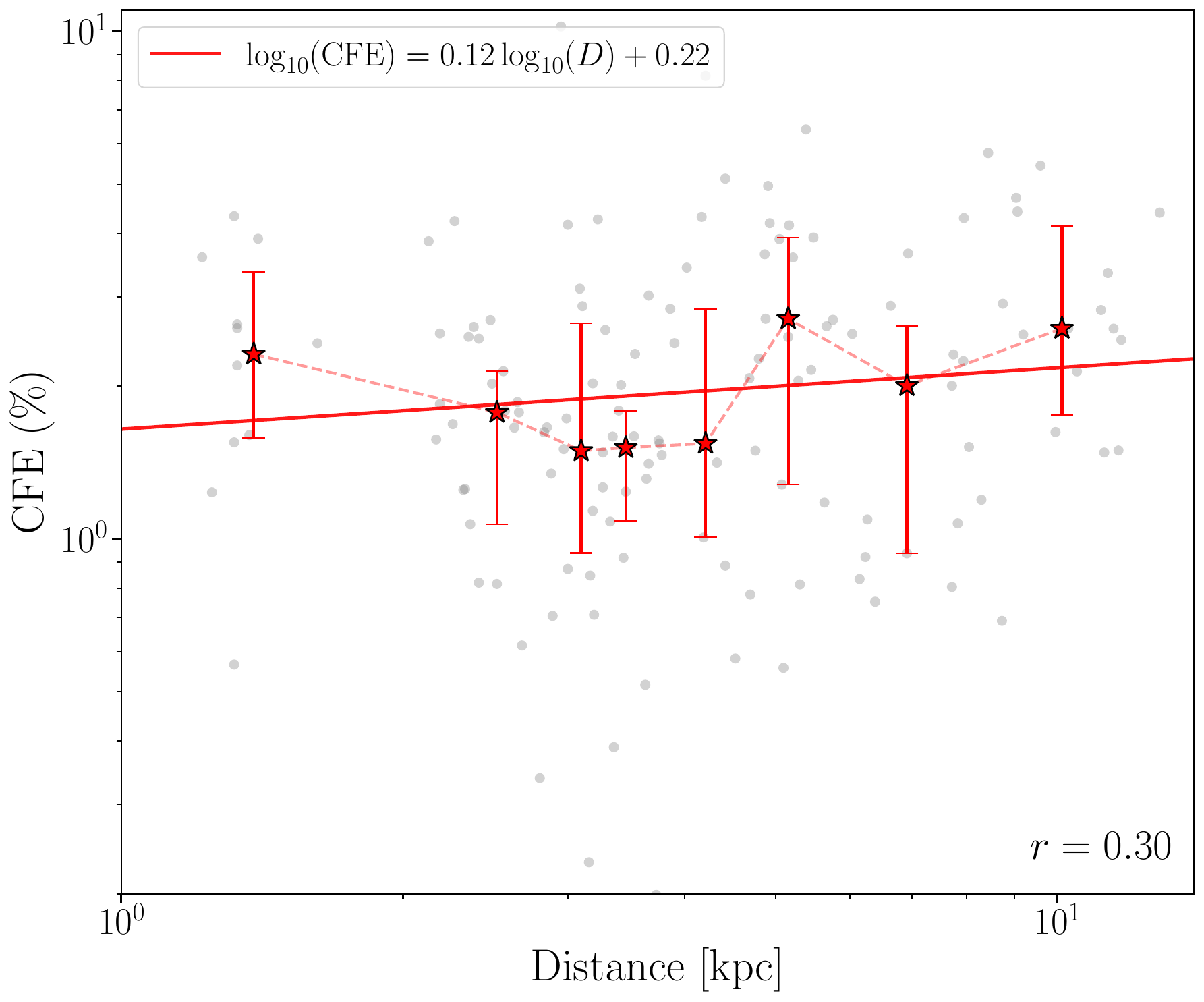}
    \vspace{-1mm}
\end{minipage}

\caption{Same as Fig.~\ref{fig:cfe_lm}, but for the correlations between CFE and the clump-averaged density (left) and source distance (right).}
\label{fig:CFE_VS_parameter}

\end{figure*}

In the main text, we show the correlation between the CFE and the clump luminosity-to-mass ratio ($L/M$). Here we further present the correlations between the CFE and two additional parameters: the clump-averaged density and the source distance. Fig.~\ref{fig:CFE_VS_parameter} illustrates these relationships, and no significant correlation is found.


%


\bibliography{sample701}{}
\bibliographystyle{aasjournalv7}



\end{document}